\documentclass[%
 aip,
 amsmath,amssymb,
preprint,%
]{revtex4-1}

\DeclareSymbolFont{matha}{OML}{txmi}{m}{it}
\DeclareMathSymbol{\varv}{\mathord}{matha}{118}
\usepackage[normalem]{ulem}

\usepackage{graphicx}
\usepackage{float}
\usepackage{dcolumn}
\usepackage{bm}

\usepackage[utf8]{inputenc}
\usepackage[T1]{fontenc}
\usepackage{mathptmx}
\usepackage{etoolbox}

\usepackage[T1]{fontenc}
\usepackage[utf8]{inputenc}
\usepackage[absolute,overlay]{textpos}
\usepackage{hyperref}
\usepackage{graphicx}
\usepackage{dcolumn}
\usepackage{xcolor}
\usepackage{siunitx}
\usepackage[all]{nowidow}
\usepackage{subfig}

\usepackage{commath} 
\usepackage{braket}  
\usepackage{lipsum} 
\usepackage{import}
\usepackage{color}

\newcommand*{\citen}{}
\DeclareRobustCommand*{\citen}[1]{%
  \begingroup
    \romannumeral-`\x 
    \setcitestyle{numbers}%
    \cite{#1}%
  \endgroup
}

\makeatletter
\def\@email#1#2{%
 \endgroup
 \patchcmd{\titleblock@produce}
  {\frontmatter@RRAPformat}
  {\frontmatter@RRAPformat{\produce@RRAP{*#1\href{mailto:#2}{#2}}}\frontmatter@RRAPformat}
  {}{}
}%
\makeatother

\begin{document}


\title[]{Oscillations of a Water Droplet on a Horizontally Vibrating Substrate}
\author{King L. Ng*}
\email{klng@ifpan.edu.pl}
\affiliation{Institute of Physics, Polish Academy of Sciences, Al. Lotnik\'ow 32/46, 02-668 Warsaw, Poland}
\author{Lu\'is H. Carnevale}
\affiliation{Institute of Physics, Polish Academy of Sciences, Al. Lotnik\'ow 32/46, 02-668 Warsaw, Poland}
\author{Michał Klamka}
\affiliation{Institute of Aeronautics and Applied Mechanics, Warsaw University of Technology, Nowowiejska 24, 00-665 Warsaw, Poland}
\author{Piotr Deuar}%
\affiliation{Institute of Physics, Polish Academy of Sciences, Al. Lotnik\'ow 32/46, 02-668 Warsaw, Poland}
\author{Tomasz Bobinski}
\affiliation{%
Institute of Aeronautics and Applied Mechanics, Warsaw University of Technology, Nowowiejska 24, 00-665 Warsaw, Poland
}%
\author{Panagiotis E. Theodorakis}%
\affiliation{Institute of Physics, Polish Academy of Sciences, Al. Lotnik\'ow 32/46, 02-668 Warsaw, Poland}

\date{\today}

\begin{abstract}
Deformed droplets are ubiquitous in various industrial applications,
such as inkjet printing, lab-on-a-chip devices, and spray cooling, and can fundamentally affect the involved applications both
favorably and unfavorably.  Here, we employ
many-body dissipative particle dynamics to investigate
the oscillations of water droplets on a harmonically and 
horizontally vibrating, solid substrate.
Three distinct scenarios of oscillations as a response to the substrate
vibrations have been identified. The first scenario reflects a common
situation where the droplet can follow the 
substrate vibrations. In the other two scenarios, favored in the case of hydrophilic substrates, droplet 
oscillations generate high shear rates that ultimately lead to 
droplet breakup. Leveraging our simulation model, the properties 
of the droplet and the mechanisms related to the oscillations are
analyzed with a molecular-level resolution, while
results are also put in the perspective of experiment. Our study suggests that the three scenarios can be
distinguished by the contact-surface velocity of the oscillating
droplet, with threshold velocities influenced by the substrate’s wettability.
Moreover, the mean magnitude of the particle velocity at
the contact surface plays a key role in determining the three 
oscillation phases, suggesting that the capillary number of the oscillating droplet governs the phase behavior.
Thus, our approach aims to optimize droplet oscillations and deformations on solid substrates,
which have direct implications for technological applications.
\end{abstract}

\maketitle

\section{Introduction}

Many industrial applications involve deforming droplets on a
solid substrate, for example, in 
lab-on-a-chip 
applications,\cite{Dkhar2023}
inkjet printing,\cite{Hoath2016} spray cooling,\cite{Yin2022} 
and combustion,\cite{Kummitha2024} just to mention a few. In such
applications, vibrations can impact the performance at the elementary level.\cite{Timonen2013} For example, on the one hand,
droplet shape oscillations caused by vibrations can lead to
undesirable droplet breakup\cite{carnevale2023}.
On the other hand, understanding the role of oscillations at the
droplet level can be exploited to enhance the mixing
within the droplet in microfluidics applications\cite{Mugele2006,Daniel2005} or in an oscillating droplet
tribometer.\cite{Junaid2022}

The oscillations of drops, in general, have been investigated in the
case of free
droplets, namely those falling or floating freely in their surroundings,
as well as constrained drops, for example, a drop constrained
by a rod
that is forced to oscillate in a fluid of similar
density to remove the effects of gravity, thus enabling the study
of pure surface tension restored oscillations.\cite{Rodot1979}
Moreover, oscillations can be categorized as
axisymmetric or non-axisymmetric surface oscillations,
that is, surface oscillations caused by vertical or 
horizontal vibrations, respectively.\cite{Milne2014} 
While the oscillation of constrained drops has only
been considered over the last decades,\cite{Rodot1979} that 
of free drops is a much more 
long-standing topic of the literature.
For example, Lord Rayleigh has proposed a formula for the
frequency of axisymmetric capillary wave oscillations of 
spherical
drops through the balance of surface tension 
and inertial forces, almost two centuries 
ago\cite{Rayleigh1945} (for a discussion on Rayleigh
modes, free and constrained, see Refs~\citen{Daniel2005,book:chandrasekhar,Mettu2008,Lyubimov2006}),
while Lord Kelvin\cite{Kelvin1890} and Lamb\cite{Lamb1881} 
considered oscillations balancing
inertia and gravity. The focus of those investigations
has been to theoretically
describe the ordinary frequency of surface oscillation
modes in various frameworks, which can be classified based
on the restoring force for oscillations, such as surface tension or gravity. 
A closely related concept is that of droplets on inclined 
substrates,\cite{Pierce2008,Brunet2007,Huber2019}
where it has been demonstrated
that liquid drops on an inclined plane, oscillating 
vertically are even able to climb 
uphill.\cite{Benilov2011,Brunet2007}

Despite a persistent interest in droplets on vibrating
substrates and their technological implications, there
are only a handful of studies in this area. They have 
employed experimental and theoretical methods to mainly
describe the various modes of droplet 
surface-oscillations.\cite{Becker1991,Becker1994,Rahimzadeh2019,Pyankova2022,Dong2006,Celestini2006,Ding2010,Blake2006,Daniel2002,Zhang2024,Deepu2014,Gilewicz_Master_Thesis,Sharp2011,Sharp2012}
In particular,
Becker \textit{et al.} \cite{Becker1991,Becker1994}
have highlighted the
role of nonlinear effects, for example, concerning
the dependence of
the oscillation frequency on the amplitude, as well as
an asymmetry of the oscillation amplitude and interaction
between the different oscillatory modes. Their
experiments and theoretical analysis have indicated
that the full theoretical description of
droplet oscillations is much more complex than
suggested by the analysis above,\cite{Milne2014} 
and that linear approximations
are only relevant for small amplitude due to
viscous damping. The nonlinear frequency response
of oscillations of sessile droplets subjected to horizontal
vibrations has also been the topic of a more recent 
study.\cite{Rahimzadeh2019} By means of experiments,
it has been found that the droplet oscillates at a response 
frequency a multiple of the excitation frequency, but, only,
for a sufficiently high vibration amplitude and a lower
excitation frequency than the natural frequency of 
the sessile droplet. Moreover, at higher excitation
frequency, the droplet oscillates at around the
excitation frequency,
while at ultrasonic excitation frequency, the droplet
cannot follow the excitation.
There is therefore a physical limitation
for forming infinite modes of infinite wavenumber on the 
surface of a small droplet.

Besides those investigations, further understanding of droplet
oscillations on vibrating substrates still remains elusive
in the literature.
In the case of experiments, this is mainly due to the fact that 
experimental studies face limitations when
attempting to capture the
various modes by tracking the surface oscillations of
the drops or when trying to obtain information on the
velocity patterns within the droplets.
This practically means that there is a limited
number of possible experimental pathways that can be taken
to investigate these dynamic phenomena. One must assume
that patterns will remain periodic, while, at the same time, data
can depend strongly on a whole set of parameters
(such as substrate roughness and evaporation),
that are generally difficult to control and may vary between
different experiments. With most 
research remaining focused on identifying the
modes of the oscillating droplet surface,
even the characterization of the natural modes has thus
far had an incomplete degree of success.\cite{Strani_Sabetta_1984,Chiba2012,Lyubimov2006,Noblin2004,Sharp2011,Sharp2012}
For example, the concept of 
coinciding frequency with one of the natural frequencies
has theoretically been used to model 
droplet oscillations.\cite{Kalmar-Nagy2011,Chen2016,Rahimzadeh2019,Mettu2008,Celestini2006,Deepu2014,Lin2018,Li2024}
In this case, a nonlinear mass--spring-damper, 
known as a Duffing oscillator,\cite{Kalmar-Nagy2011} has usually 
been employed to 
investigate axisymmetric modes of a sessile droplet
due to vertical vibrations.\cite{Deepu2014}
In other words, theory has predominantly been
based on a mathematical
analysis that includes various model assumptions, 
such as the type of restoring force, without
taking into account the complex nature of oscillation
phenomena, such as the flow within the droplets, the
liquid--vapor coexistence, evaporation and physical 
pinning. Many of these assumptions are highly nonlinear, and despite
efforts,\cite{Rahimzadeh2019}
disentangling the interplay of the various parameters that determine the droplet behavior
still remains a challenge.
These experimental and theoretical
limitations therefore point to evident
gaps in our understanding of a fundamental phenomenon
in nature. One that, moreover, is highly relevant
for practical applications.

In view of these limitations, a possible way forward concerning
the investigation of non-axisymmetric 
droplet oscillations on
vibrating substrates would be the use of molecular-level methods
based on suitable force-fields, such as
molecular dynamics.\cite{Rapaport_2004}
In this case, there is a need to reach an adequately 
large droplet size, such to allow for clear
oscillation observations, flow fields, and possibly
phase transitions, which naturally 
renders coarse-grained models
the `go-to' option.\cite{Theodorakis2015Langmuir}
However, again, simulating droplets of that size and carrying out a
full investigation of droplet oscillation phenomena 
directly via this method is
computationally costly, if not prohibitive, since it
entails the investigation of a range of different droplet sizes
to assess finite-size effects, as well as
the exploration of parameters
such as the frequency and amplitude of the oscillations
for various substrate-wettability scenarios. 
Moreover, such models generally do not
capture well some hydrodynamic effects or lack precise 
predictive power for some of the liquid properties, 
such as dynamic viscosity. These properties
are required to be described correctly
due to the competition
of viscous and surface-tension forces, where the latter
can act as the restoring force of the droplet shape.

Therefore, we have chosen to use
the method of many-body dissipative particle dynamics (MDPD)
in this study, which is a method
that can easily handle large droplets and, moreover,
can take into account hydrodynamic 
effects.\cite{Carnevale2024,carnevale2023}
At the same time, suitable interactions can be obtained for a range
of different compounds, including complex fluids 
beyond pure water droplets, which will specifically be investigated 
in this work.\cite{Carnevale2024_MDPD_MARTINI}
In addition, the method includes thermal fluctuations and
as a particle-based method 
provides the necessary molecular resolution for acquiring
much needed information on both structural and dynamic
properties of the droplets, while viscosity and surface
tension are accurately reproduced by the model for
the liquid of choice. In this context, MDPD has 
been used to successfully investigate
the breakup of various liquids with a 
technologically relevant spectrum of Ohnesorge 
numbers (ratio of viscous to surface tension forces).
The breakup is caused by surface-tension 
instabilities initiated by thermal 
fluctuations \cite{carnevale2023}. 
Importantly, the MDPD  method does not require any 
\textit{ad hoc} model assumptions, 
such as a contact-angle model, in the study
of droplet oscillations,
which could often be the case in continuum models.

Hence, by harvesting the advantages of the MDPD method, 
we have embarked here on investigating oscillating water droplets
on horizontally vibrating solid substrates.
A relevant range of simulation parameters has been considered
for the amplitude and frequency of those vibrations while
taking into account droplets of various sizes.
In the simulations, we were able to
characterize in detail both the properties of droplets
for the static case, as well as the properties during substrate vibrations. 
The static case is without substrate vibrations, and
is used as a reference, for example, to determine the
natural frequency of the droplets and the equilibrium
contact angles.
Here, one of the advantages of
our method is the tight control of the experimental
conditions during the simulation and the ability to
track the instantaneous velocity and position of each particle
at every simulation time-step. In this way, properties such as
the shape of the droplet, its natural frequency,
the receding and advancing contact angles, and especially
the flow field during oscillations
can be obtained as a function of time
and with a molecular-level resolution. 
A key outcome of our research is the classification of the
various scenarios of the droplets' behavior, identifying whether 
they lead to droplet breakup, together with an unraveling of
the mechanisms associated with each
of them. This internal mechanics has largely remained unknown, 
due to the limitations of previous methods.  
With regard to internal mechanisms, a parallelism with droplet 
breakup phenomena might be of relevance.\cite{carnevale2023}

In the following section, 
we provide details on our model and methods
used in our investigations. Then, 
we present our results for both static and oscillating droplets
in Section~\ref{results},
while we draw our conclusions and
pave the way for future work in this area in Section~\ref{conclusions}.

\section{Model and Methodology}
\label{model}

Many-body dissipative particle dynamics
is a particle-based simulation method that evolved from 
dissipative particle dynamics \cite{hoogerbrugge1992,Lavagnini2021}
and it incorporates
both attractive and repulsive interactions.
This enables the simulation of fluids with free surfaces, 
making it particularly well-suited for modeling
liquid--vapor interfaces, \cite{pagonabarraga2000,warren2003}
such as liquid droplets surrounded by vapor.
In the case of MDPD, the equation of motion (Eq.~\ref{eq1}) is integrated
for each particle, $i$, which 
interacts with other particles, $j$, by means of 
a conservative force, $\bm{F}^C$, 
a random force, $\bm{F}^R$, and a dissipative force, $\bm{F}^D$.
In our work, the integration of the equations of motion is realized via
a modified velocity-Verlet algorithm \cite{groot1997}
with a time-step, $\Delta t = 0.005$ (MDPD units),
as implemented in LAMMPS software \cite{LAMMPS}, with
the equation reading
\begin{eqnarray}
m\frac{d\bm{v}_i}{dt} = \sum_{j\neq i} \bm{F}_{ij}^C + \bm{F}_{ij}^R + \bm{F}_{ij}^D.
\label{eq1}
\end{eqnarray}
The mathematical expressions of the conservative force reads
\begin{eqnarray}
\bm{F}^C_{ij} =  A\omega_{c}(r_{ij})\bm{e}_{ij} + 
B \left(\bar{\rho}_i + \bar{\rho}_j \right) \omega_{d}(r_{ij})\bm{e}_{ij}.
\label{eq2}
\end{eqnarray}
Here, $A<0$ is the attractive strength of the force,
while $B>0$ is a 
repulsive interaction strength.
The repulsive force also depends on the local density and
includes many-body effects.
The repulsive parameter, $B$, has
to be the same for all the interactions due to a known
`no-go' theorem \cite{warren2013},
otherwise the force wouldn't be conservative.
$r_{ij}$ is the distance between particles,
while $\bm{e}_{ij}$ is 
the direction vector from particle \textit{i} to particle 
\textit{j}. $\omega_{c}(r_{ij})$ and $\omega_{d}(r_{ij})$ are weight
functions that are defined as follows:
\begin{eqnarray}
\omega_{c}(r_{ij}) = 
\begin{cases}
&1 - \frac{r_{ij}}{r_{c}}, \ \ r_{ij} \leq r_{c} \\
& 0,  \  \ r_{ij} > r_{c},
\end{cases} 
\label{eq3}
\end{eqnarray}
\begin{eqnarray}
\omega_{d}(r_{ij}) = 
\begin{cases}
&1 - \frac{r_{ij}}{r_{d}}, \ \ r_{ij} \leq r_{d} \\
& 0,  \  \ r_{ij} > r_{d}.
\end{cases} 
\label{eq3.2}
\end{eqnarray}
Here, the cutoff distance for the attractive interaction is denoted
with $r_c$ and is set to unity. The repulsive-interaction cutoff
is $r_d=0.75r_c$. \cite{Vanya2018} 
The many-body contributions in the repulsive force
expressed through the local densities, $\bar{\rho}_i$ and $\bar{\rho}_j$, are calculated as follows:
\begin{eqnarray}
\bar{\rho}_i = \sum_{j\neq i} \frac{15}{2\pi r_d^3} \left( 1 - \frac{r_{ij}}{r_d}\right)^2.
\label{eq4}
\end{eqnarray}
Thermal fluctuations are taken into account through random
and dissipative forces in the equation of motion (Eq.~\ref{eq1}),
which act as the thermostat of the particles (the temperature
is $T=1$ in MDPD units for static droplets). These are 
expressed as 
\begin{eqnarray}
\bm{F}^{D}_{ij} = -\gamma_D \omega_{D}(r_{ij}) (\bm{e}_{ij} \cdot  \bm{v}_{ij})\bm{e}_{ij} ,
\label{eq5}
\end{eqnarray}
\begin{eqnarray}
\bm{F}^R_{ij} = \sigma_R \omega_{R}(r_{ij}) \xi_{ij} \Delta t^{-1/2}\bm{e}_{ij}.
\label{eq26}
\end{eqnarray}
Here, $\gamma_D$ is the dissipative strength, $\sigma_R$ the strength of the random force, 
$\bm{v}_{ij}$  the relative velocity between particles, and $\xi_{ij}$ a random 
variable from a Gaussian distribution with zero mean and unit variance. In addition,
$\gamma_D$ and $\sigma_R$ are related via the fluctuation--dissipation
theorem,\cite{espanol1995} in this case 
\begin{eqnarray}
\gamma_{D} = \frac{\sigma_{R} ^2}{2 k_B T}, 
\label{eq7}
\end{eqnarray}
while the weight functions for the forces are
\begin{eqnarray}
\omega_{D}(r_{ij}) = \left[\omega_{R}(r_{ij})\right]^2 = \left( 1 - \frac{r_{ij}}{r_c}\right)^2.
\label{eq8}
\end{eqnarray}

To simulate water droplets,  
the commonly adopted values for water in MDPD have been used,
\textit{i.e.} $A = -40$ and $B=25$
(MDPD units),\cite{zhou2019,Carnevale2024}
while the dissipative coefficient is
$\gamma_D=4.5$.\cite{Ghoufi2011, arienti2011}
The conversions from reduced to real units for the particle properties 
based on this model are reported in Table~\ref{tab:units}.


\begin{table}[b]
\caption{\label{tab:units} Relation between MDPD units and real units. The scaling 
is done by matching surface tension $\gamma$ and density of water $\rho$ to values measured from MDPD
simulations using $A=-40$ and $B=25$. The coarse-graining level is defined so that
one MDPD particle represents three water molecules. }
\begin{ruledtabular}
\begin{tabular}{lll}
Parameter  & MDPD value  & Real value \vspace{.1cm} \\ 
 \hline 
Particle &  1    & 3 H$_2$O  \\ 
$r_c$    &  1     &  8.17 \AA          \\ 
$\rho$   &  6.05  &  997 kg/m$^3$       \\
$\gamma$ &  7.62  &  72 mN/m     \\
\end{tabular}
\end{ruledtabular}
\end{table}

\begin{figure}[htb]
\includegraphics[width=0.6\columnwidth]{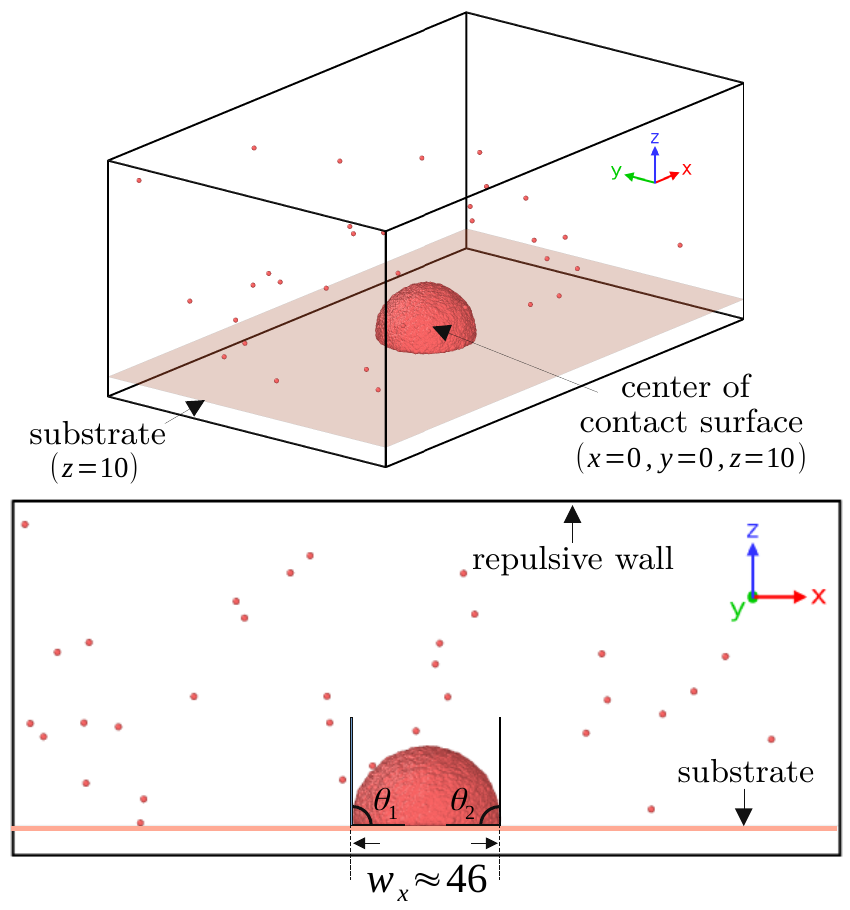}
\caption{A typical snapshot in the simulations of a water 
droplet on a static substrate. In the lower panel, 
$\theta_{1}$ and $\theta_{2}$ indicate the contact
angles of the droplet at the two points along
the contact line, as indicated. Here, the number of particles 
$N=20\times10^4$ and the droplet--substrate affinity 
$\varepsilon_{\rm ws}=2.0$, resulting in equilibrium contact
angles $\theta_1,\theta_2\approx90^{\circ}$. 
The droplet width at the static contact surface is 
$w_{x}\approx46$. 
Snapshots were generated using OVITO software.\cite{Stukowski2010}
}
\label{fig:static_dpl_ovito}
\end{figure}

\begin{figure}[htb]
\includegraphics[width=0.6\columnwidth]{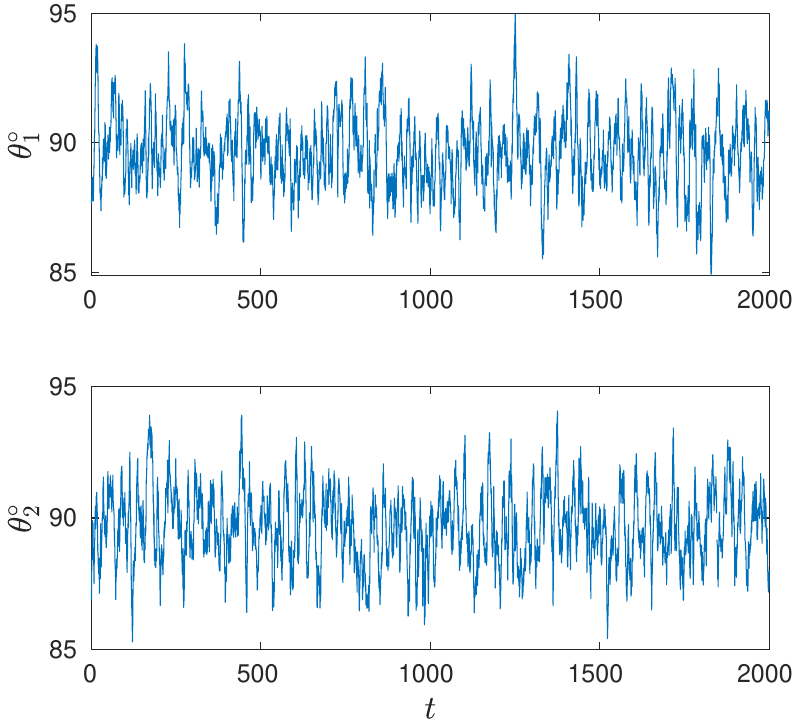}
\caption{Contact angles $\theta_{1}$ and $\theta_{2}$ of a static droplet over simulation time, $t$. Same parameters as in Figure~\ref{fig:static_dpl_ovito} where $N=20\times10^4$ and $\varepsilon_{\rm ws}=2.0$. The values shown above
are averages from four samples. The contact angles are about $90^{\circ}$.}
\label{fig:static_contact_angle}
\end{figure}

To prepare the simulation for investigating the droplet oscillations
(both natural ones and those caused by the vibrating substrate),
freely suspended droplets of different sizes were first equilibrated
in a simulation box with dimensions
$L_x = 1200$, $L_y= 220$, and $L_z= 120$ (MDPD units). The total
number of particles in the simulation were 
$N=5\times10^4$, $N=10\times10^4$, or $N=20\times10^4$ particles, 
which correspond to spherical droplets with radii
$R=(\frac{3N}{4\pi\rho})^{1/3}=12.54,15.80$ and $19.91$ (MDPD units), respectively. 
For the larger droplets with $N \geq 10\times10^4$, 
key properties for the droplets, such as the contact angle of the 
droplet, seem not to depend on the droplet 
size.\cite{theodorakis2015modelling}
To maintain the same thermodynamic conditions 
given the number of particles and the temperature of the system, 
we keep the volume available to the water particles the same
by means of implicit walls, with further details on these walls
given later.
Each droplet, which consists of slightly less than $N$ particles due
to evaporation,  was placed in the middle of the simulation box at
position ($x=0$, $y=0$)
and on a smooth, unstructured, solid
substrate (at $z=10$) (Figure~\ref{fig:static_dpl_ovito}). 
This substrate was implicitly modeled via the Lennard-Jones (LJ)
potential,\cite{Theodorakis_MCVOF_2021} mathematically
expressed as follows:
\begin{equation}\label{eq:LJpotential93}
U^{\rm 9-3}(z') = 4\varepsilon_{\rm ws} \left[  \left(\frac{\sigma_{\rm  ws}}{z'}
\right)^{9} - \left(\frac{\sigma_{\rm ws}}{z'}  \right)^{3}    \right].
\end{equation}
Hence, the force included in the equations of motion for
each particle that interacts with the substrate when 
its distance is below the LJ cutoff, $r_{\rm LJ}=1.0$, is obtained to be $F_{\rm ws}=-\frac{\partial U^{\rm 9-3}(z')}{\partial z'}$.
Here, $z'$ is the distance between the liquid particles
and the substrate, $\sigma_{\rm ws}$ is set to unity, 
while $\varepsilon_{\rm ws}$ is a parameter used to tune the
affinity of the droplet to the substrate, thus controlling
the equilibrium contact angle of the droplet. The
values  $\varepsilon_{\rm ws}$ chosen were
droplet-size-dependent.
For $N=20\times10^4$, the exact values of the interaction
strength used were
$\varepsilon_{\rm ws}$ = 3.0, 2.5, 2.0, 1.0, and 0.1, 
which correspond to contact angles
$\theta=50^\circ$, 65$^\circ$, 90$^\circ$, 115$^\circ$, 
and 140$^\circ$, respectively. For $N=10\times10^4$,  
$\varepsilon_{\rm ws}$ = 3.5, 3.0, 2.0, 1.0, 0.1 
for the same corresponding angles; 
for $N=5\times10^4$,  
$\varepsilon_{\rm ws}$ = 3.5, 3.0, 2.0, 1.0, 0.25, respectively.
Overall, $\varepsilon_{\rm ws}$ was chosen such to produce the
same equilibrium contact-angles for all sizes of droplets.
This range of contact-angles cover
hydrophilic and hydrophobic substrate cases, which
are relevant experimentally.\cite{Gilewicz_Master_Thesis}
A perspective and orthogonal view of a sessile droplet
for the static substrate case with equilibrium
contact angle 90$^\circ$ are presented in
Figure~\ref{fig:static_dpl_ovito}, while typical
measurements of the contact angle are presented in
Figure~\ref{fig:static_contact_angle}. In the case shown,
the standard deviation of the contact angle measurements
in a single trajectory are of the order of 2$^\circ$.
Finally, a purely repulsive LJ wall was used to prevent particles
from escaping from the top of the simulation box 
and sticking to the bottom
substrate from below, due to the presence of
periodic boundary conditions on all sides of 
the simulation box.

To determine the contact angle,
we use a polynomial method up to order $n$ 
to obtain the surface curvature of the $y=0$ cross-section
of the droplet (like the lower panel of Figure~\ref{fig:static_dpl_ovito}), given by
\begin{equation}\label{eq:polynomial}
x = p(z) = p_{1}z^{n}+p_{2}z^{n-1} + \ldots + p_{n}z+p_{n+1},
\end{equation}
where $p_{n}$ are the coefficients obtained from least-squares 
fitting. In the simulations, values of $3\leq n\leq5$ best fit the surface 
profile. The contact angle is then determined from the tangent line of 
the polynomial function (Equation~\ref{eq:polynomial}) at the droplet 
contact-line as shown in Figure~\ref{fig:static_dpl_ovito}. 
Since the stationary droplet is generally axisymmetric 
along the $z$ direction,  two
contact angles were measured along the $x$
direction (on the $x-z$ plane) to improve statistical accuracy for the stationary droplet instead of one required
based on a sufficient number of statistically independent
ensemble droplet-snapshots. We will later acquire separate
measurements of the advancing and receding contact angles
when substrate vibrations in the $x$ direction are
activated. In this work, we use the notation $\theta_1$ and
$\theta_2$ instead of the more
common notation $\theta_{\rm r}$ and $\theta_{\rm a}$ for
the receding and advancing contact-angles, respectively.
This choice reflects the alternating nature of the angles for oscillating
droplets on a vibrating substrate. Thus, the left
angle (toward the negative $x$ axis)
will always be denoted as $\theta_1$, while the
right angle (in the positive $x$ direction)
will be denoted as $\theta_2$,  
as shown in Figure~\ref{fig:static_dpl_ovito}.

To induce the droplet oscillations, the substrate is vibrated
non-axisymmetrically, in our case in the $x$ direction. 
The characteristics of the substrate vibration, 
\textit{i.e.} the amplitude and the frequency of the 
oscillation, are set as inputs to the simulation
and remain constant throughout.
The instantaneous velocity of the
substrate, is given by:
\begin{equation}\label{eq:substrate_velocity}
u_{\rm{sub}}(t) = -A_{\mathrm{sub}}\omega_{\mathrm{sub}}\mathrm{sin}(\omega_{\mathrm{sub}}t).
\end{equation}
$A_{\mathrm{sub}}$ is the vibration amplitude, and $\omega_{\mathrm{sub}}=2\pi f_{\mathrm{sub}}$ 
the angular frequency, with $f_{\mathrm{sub}}$ 
the vibration frequency. 
The range of parameters used for the substrate vibrations, has been chosen to be in the vicinity of the natural oscillation frequency
of the droplets determined from
our simulations and experiments,\cite{Gilewicz_Master_Thesis} 
specifically $225\leq A_{\mathrm{sub}}\leq900$ and 
$3.75\times10^{-3}\leq f_{\mathrm{sub}}\leq15\times10^{-3}$ (MDPD units).

It is important to relate our explicit $\sim10$nm radius droplets 
to droplets with macroscopically observable size. 
This can be carried out since it is known that the larger
droplets are already in the appropriate scalable 
regime.\cite{theodorakis2015modelling}
The characteristic inertial timescale of a droplet can be estimated
as $t_{c}=\sqrt{\frac{\rho R^3}{\gamma}}$.
For the droplet with $N=20\times10^4$ and $R=19.91$, 
this yields $t_{c}=79.16$ (MDPD unit). 
For a macroscopic droplet with a 
radius of $R^{\rm real}=1\rm{mm}$ 
(where the superscript "$\rm real$" denotes real values in SI units 
of droplets scaled to macroscopic sizes), 
we obtain $t^{\rm real}_{c}=\sqrt{\frac{\rho^{\rm real}(R^{\rm{real}})^3}{\gamma^{\rm real}}}=3.72\times10^{-3}\rm{s}$.
The oscillation frequency of the substrate in ``real'' macroscopic 
droplet units can then be estimated, where
$f^{\rm real}_{\rm sub}=f_{\rm sub}t_{c}/t^{\rm real}_{c}$. 
For example, for a droplet with a radius of $1\rm mm$, 
the corresponding frequency is approximately
$80{\rm Hz}\lesssim f^{\rm real}_{\rm sub}\lesssim 320{\rm Hz}$. 
Here $A_{\rm sub}$ is the vibration amplitude for the smooth
and unstructured surface where friction is not included in the 
simulations; hence, the real oscillation amplitude of the substrate, 
$A^{\rm real}_{\rm sub}$, is estimated based on the oscillation 
amplitude of the mean particle $x$ position 
at the contact surface, $x_{\rm cs}$. 
Using the scaling relation 
$A^{\rm real}_{\rm sub}=x_{\rm cs}R^{\rm real}/R$,
we obtain
$0.5{\rm mm}\lesssim A^{\rm real}_{\rm sub}\lesssim 4{\rm mm}$.

To conduct our investigations, a droplet is initially
equilibrated on the substrate until the standard deviation of the contact angles, $\sigma_{\theta_{1}}$ and $\sigma_{\theta_{2}}$, are of the order of $2^{\circ}$. A snapshot of the droplet at relaxation equilibrium is as shown in Figure~\ref{fig:static_dpl_ovito}. Vibrations are then activated only after the droplet has reached equilibrium.  The simulation duration varies depending on the observed droplet behavior, ensuring that the 
set of observed phenomena are not truncated by an insufficient
timescale of the simulation. Consequently, the typical simulation
time for vibrating substrates ranges from $5000 \leq t \leq 10000$ 
(MDPD units, natural for the model). We shall note
here that for the case of molecular dynamics simulations
of physics-based models based on a similar equation
of motion, the corresponding
times  in terms of natural units in the model
would be significantly longer.\cite{Theodorakis2011}
Therefore, computation time would grow accordingly.
Average properties in static cases were calculated from an 
ensemble of statistically independent configurations
(typically four), 
while dynamic properties such as flow fields
were continuously monitored
and estimated using a running average of typically
four consecutive configurations, 
obtained at a suitable short-time interval during the simulation,
namely $t_{\rm{int}}=7.5$.

\section{Results and Discussion}
\label{results}

\subsection{Droplet on a static substrate}

\captionsetup[subfigure]{labelformat=empty}
\begin{figure}[htb]

    \centering
    \subfloat[\centering ]{{\includegraphics[width=0.33\columnwidth]{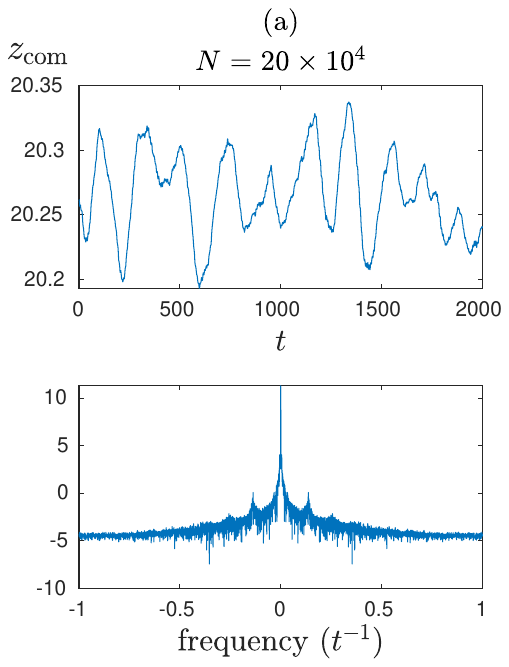} }}%
    \subfloat[\centering ]{{\includegraphics[width=0.33\columnwidth]{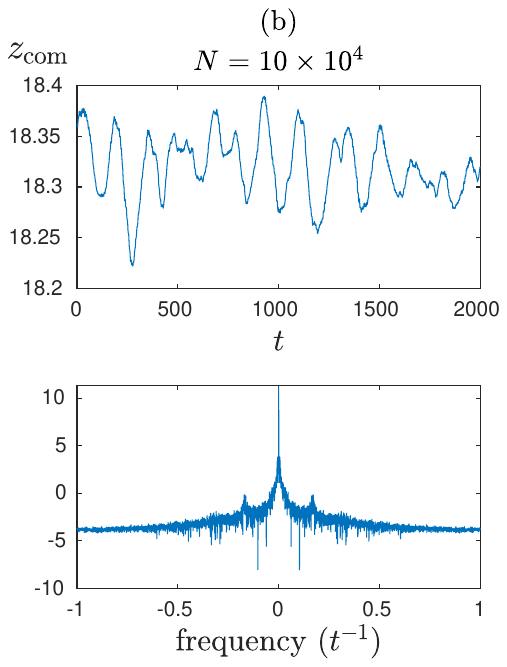} }}%
    \subfloat[\centering ]{{\includegraphics[width=0.33\columnwidth]{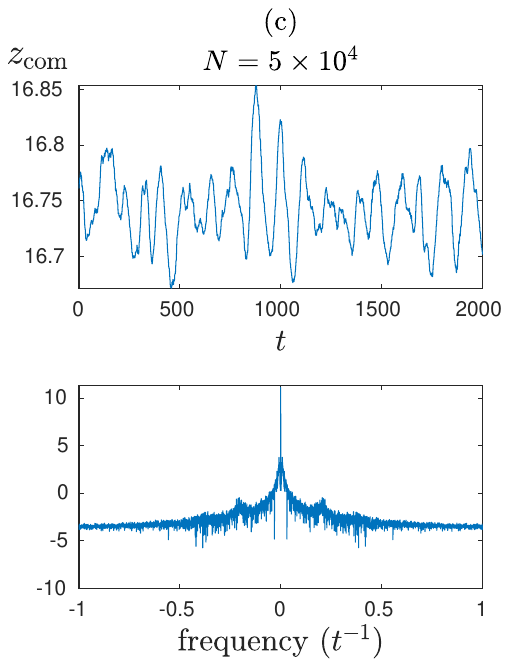} }}%

\caption{Static droplet analysis for: (a) Number of particles 
$N=20\times10^4$, (b) $N=10\times10^4$ and (c) $N=5\times10^4$,
with contact angles $\theta=90^{\circ}$.
Upper panels: The center-of-mass position ($z_{\rm{com}}$) 
of the static droplet over time $t$.
Lower panels: Fourier analysis of $z_{\rm{com}}$ 
in natural log scale vertically. 
Three eigenmodes are observed in each case (a), (b), and (c). 
For (a) the frequencies of the eigenmodes are ($t^{-1}$) 
$\sim\pm0.0052$, $\pm0.14$ and $\pm0.26$; 
(b) $\sim\pm0.0057$, $\pm0.17$ and $\pm0.32$; 
(c) $\sim\pm0.0067$, $\pm0.20$ and $\pm0.37$. 
All results are obtained from an average of four different runs.
}
\label{fig:COMz_freqz}
\end{figure}

\begin{figure}[htb]
\includegraphics[width=0.6\columnwidth]{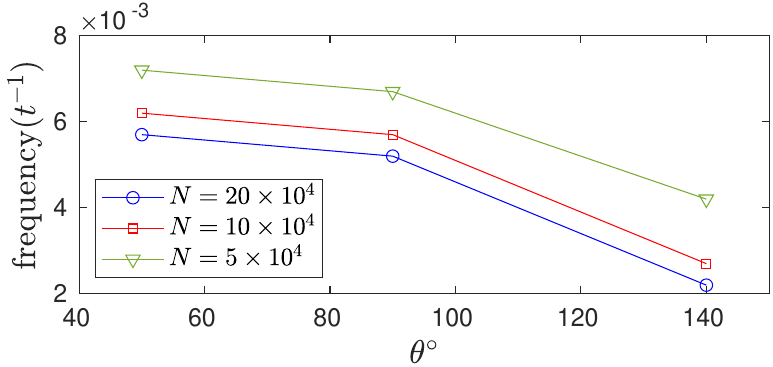}
\caption{Frequency of the $1^{\rm st}$ oscillation mode for static droplets with number of particle $N=5\times10^4$, $N=10\times10^4$ and $N=20\times10^4$, each with contact angles $\theta=50^{\circ},90^{\circ}$ and $140^{\circ}$. All results are obtained from an average of four different runs.}
\label{fig:freqz_dcomz}
\end{figure}

Droplets were investigated on static substrates to determine various reference
properties, including the wettability
of the substrate through the equilibrium contact angle.
A key property in the context of this study is the natural
eigenfrequency of the droplets, which depends on both 
the droplet size and substrate wettability.
In our case, this can be measured
not only by tracking the surface 
oscillations, but, also, oscillations of 
the center-of-mass of the droplet individually in all Cartesian
directions, as well as the collective motion of the particles
in these directions. Previous studies have employed methods 
such as principal component analysis
(also, known as principal
orthogonal decomposition) to determine the droplet 
eigenfrequencies based on droplet surface oscillations.\cite{Deepu2014}
However, these methods are computationally
expensive for droplets with tens of thousands
of particles, like what we have here.

We find that
measuring the values of $z_{\rm com}$,
the center-of-mass of the droplet 
in the $z$ (away from the substrate) direction, 
over time and applying a Fourier transform to these data 
(Figure~\ref{fig:COMz_freqz}) provides accurate results for
the main natural eigenmodes of the droplets.
The lateral droplet oscillations in the $x-y$ plane 
yield the same
results as $z_{\rm com}$, indicating
that the lateral and vertical oscillation modes of the
droplet are entirely interrelated. A similar
conclusion has been drawn when the droplet eigenfrequencies
were calculated based on all particle coordinates, albeit
noise in the data was slightly higher in this case
due to the thermal
fluctuations of the particles and the different environment
between particles in the bulk and the interfaces.

From the eigenfrequency measurements for
a droplet with a size of about $N=20\times10^4$ and a contact angle $\theta=90^{\circ}$,
we find that the $1^{\rm st}$ natural oscillation mode corresponds to a low frequency $~\pm0.0052$, and two higher characteristic 
frequencies for the $2^{\rm nd}$ and $3^{\rm rd}$ natural oscillation modes at $\pm0.14$ and 
$\pm0.26$, respectively. The $3^{\rm rd}$ weaker eigenmode occurs at 
approximately double the frequency of the
$2^{\rm nd}$ eigenmode (Figure~\ref{fig:COMz_freqz}(a)).
This indicates that nonlinear effects are indeed
present during sessile droplet oscillations, possibly due
to the presence of thermal fluctuations. 
Using the relation $f^{\rm real}=ft_{c}/t^{\rm real}_{c}$ 
for the natural frequency, $f$, and assuming a droplet with a
real spherical radius of $1\rm mm$, 
these values correspond to natural oscillation modes of 
$\sim110 \rm Hz$, $3 \rm kHz$ and $5.5 \rm kHz$.
In addition, we observe that the 
natural eigenmodes shift slightly in smaller droplets, 
with frequencies of the modes increasing moderately 
($\sim20\%$) each time 
the number of particles is reduced by half
(Figures~\ref{fig:COMz_freqz}(b) 
and \ref{fig:COMz_freqz}(c)). 
This is in line with the rule of thumb that frequency is inversely 
proportional to object size, since $(0.5)^{1/3}\approx 0.8$. 
For droplets with a contact angle of $\theta=140^{\circ}$, 
across all three sizes $N$, the frequency spectrum is dominated 
by the $1^{\rm st}$ low-frequency mode, while the $2^{\rm nd}$ 
and $3^{\rm rd}$ modes are not observed in the Fourier analysis.
We compared the $1^{\rm st}$ mode at various contact angles
$\theta$ for droplets of all three sizes 
(Figure~\ref{fig:freqz_dcomz}). 
The $1^{\rm st}$ natural oscillation mode increases non-linearly
with increasing wettability for all tested droplet sizes, 
and the frequency change of the $1^{\rm st}$ mode is 
less pronounced for hydrophilic surfaces.
Overall, the natural eigenfrequency of the droplets leads to 
a variation in $z_{\rm com}$ of approximately
$\lesssim 1\%$ across all three sizes.

\subsection{Droplet oscillation}

\begin{figure}[htb]
\includegraphics[width=\columnwidth]{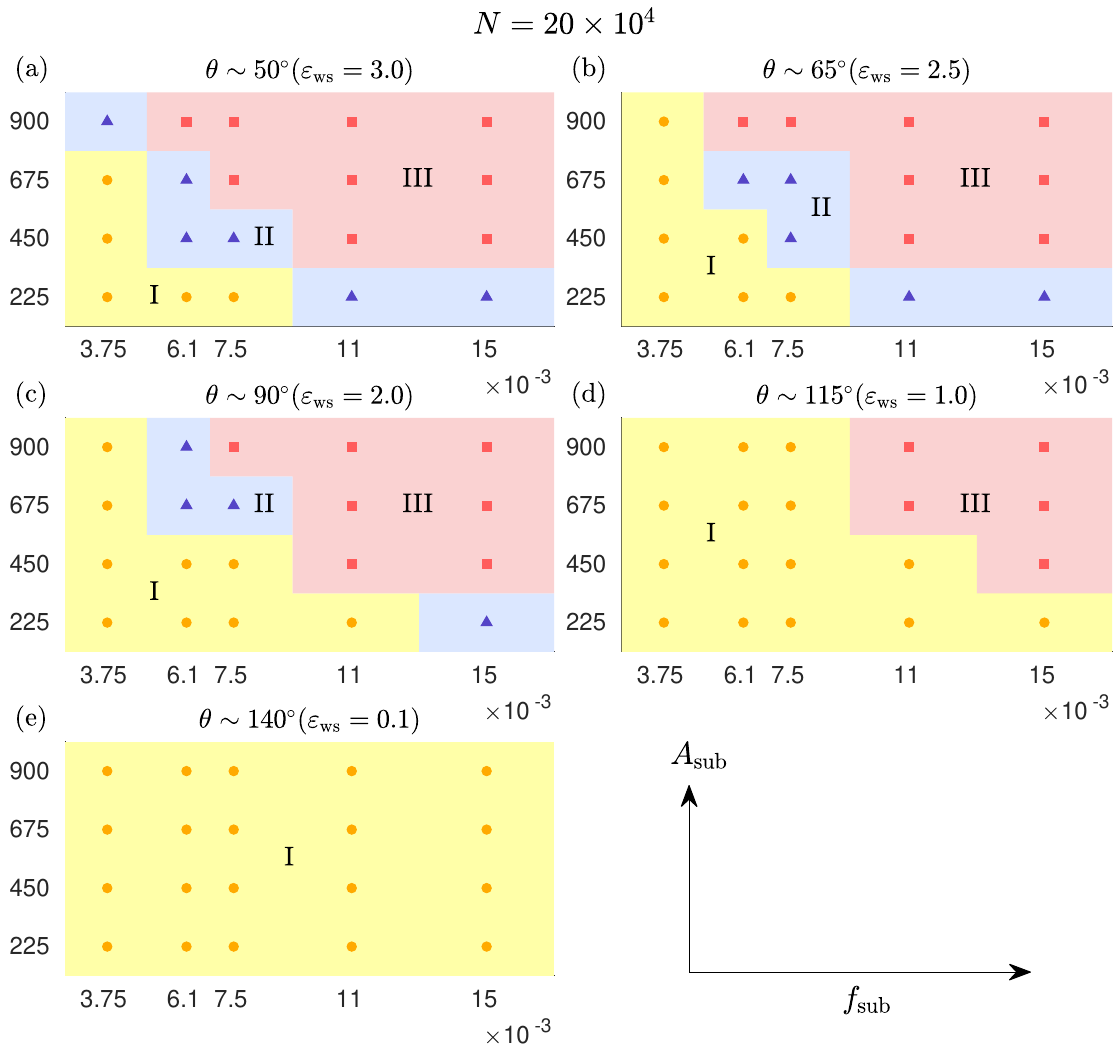}
\caption{State diagram showing Phase I, II, and III, indicated by different color shading (see text
for details), as a function of the amplitude and frequency of substrate vibrations for a droplet with 
$N=20\times10^4$. Each plot corresponds
to a substrate with different wettability, with 
the equilibrium contact angle, $\theta$, of the droplet
on a static substrate (without vibrations) indicated at the top
of each plot: (a) $\theta\sim50^{\circ}$; (b) $\theta\sim65^{\circ}$; (c) $\theta\sim90^{\circ}$; (d) $\theta\sim115^{\circ}$; (e) $\theta\sim140^{\circ}$.}
\label{fig:phase_large_droplet}
\end{figure}

Having determined the natural frequency of the droplets
across our range of droplet sizes and substrate wettability,
we took the next steps to investigate the
various scenarios unfolding for the oscillating droplets
when substrate vibrations were activated.
Our results for different amplitude and frequency
ranges are summarized in
Figure~\ref{fig:phase_large_droplet} for droplets with 
$N=20\times10^4$ MDPD particles, while results for
smaller droplets are provided in the Supplementary Material (SM). 
The behavior of the smaller droplets very much follows the same 
qualitative trends, therefore in what follows we describe the 
$N=20\times10^4$ case and concentrate on that.

The most trivial scenario occurs when the droplet exhibits
symmetric oscillations. That is, oscillations repeat periodically
with the same period as the substrate vibrations given
by Equation~\ref{eq:substrate_velocity}. 
This occurs always within our range of parameters 
for the substrate with a contact angle $\theta=140^\circ$.
In this case, the droplet remains stable with no
breakup of the droplet occurring at any time during the simulations,
which is confirmed by running multiple, different trajectories 
with different initial velocity seed over extended times. 
This scenario will be named
here as `Phase I', and the characteristics of such
oscillations will be described in detail later.

As the substrate wettability increases
(Figure~\ref{fig:phase_large_droplet}(d)), though 
the substrate still remains hydrophobic (
\textit{i.e.}, $\theta=115^\circ$), a different droplet behavior
emerges at high amplitude and frequency of substrate vibration. 
In particular, at lower oscillation amplitudes, the Phase I scenario
persists regardless of vibration frequency. 
However, as the frequency $f_{\rm sub}$ and amplitude $A_{\rm sub}$ increase, `Phase III' emerges, where the
immediate breakup of the droplet takes place as a result
of the vibrations. Details regarding the breakup will be discussed in later sections.

For substrates with an equilibrium contact angle
$\theta=90^\circ$, `Phase II' scenario appears in our simulations at high oscillation frequency and amplitudes
(Figure~\ref{fig:phase_large_droplet}(c)). In Phase II, the droplet exhibits symmetric oscillations
for a prolonged time before an asymmetry in the oscillations
arises, leading to the breakup of the droplet.
We observe that Phases II and III
are more prevalent than Phase I droplet oscillations
for the range of frequencies and amplitudes considered
here. In particular, Phase III again emerges at higher oscillation amplitudes and frequencies, while Phase II appears along the boundary between
Phase I and III. When the equilibrium contact angle
becomes less than 90$^\circ$, Phase III occurs at even
lower frequencies and 
amplitudes (Figure~\ref{fig:phase_large_droplet}(a) and (b)).
However, there appear to be threshold
values for both frequency and amplitude even at smaller
contact angles. In particular, Phase I is observed consistently in 
all cases when the frequency is below $7.5 \times 10^{-3}$
and amplitude below $675$. 
The state diagrams for smaller droplets
presented in Supplementary Material reveal similar trends, 
with Phase II appearing more prominently along the boundary 
between Phase I and III, and conditions for droplet breakup being
more favorable.
This is an indication that surface-tension forces,
which are more dominant in smaller droplets
due to increased surface-area-to-volume interactions
would promote the droplet breakup.
Overall, the breakup of the droplets
is favored by higher amplitudes and frequencies,
especially for droplets on  hydrophilic substrates with
smaller contact angles. Details for the characteristics of both Phase II and Phase III
will be discussed in the following subsections.

\begin{figure}[htb]
\includegraphics[width=0.5\columnwidth]{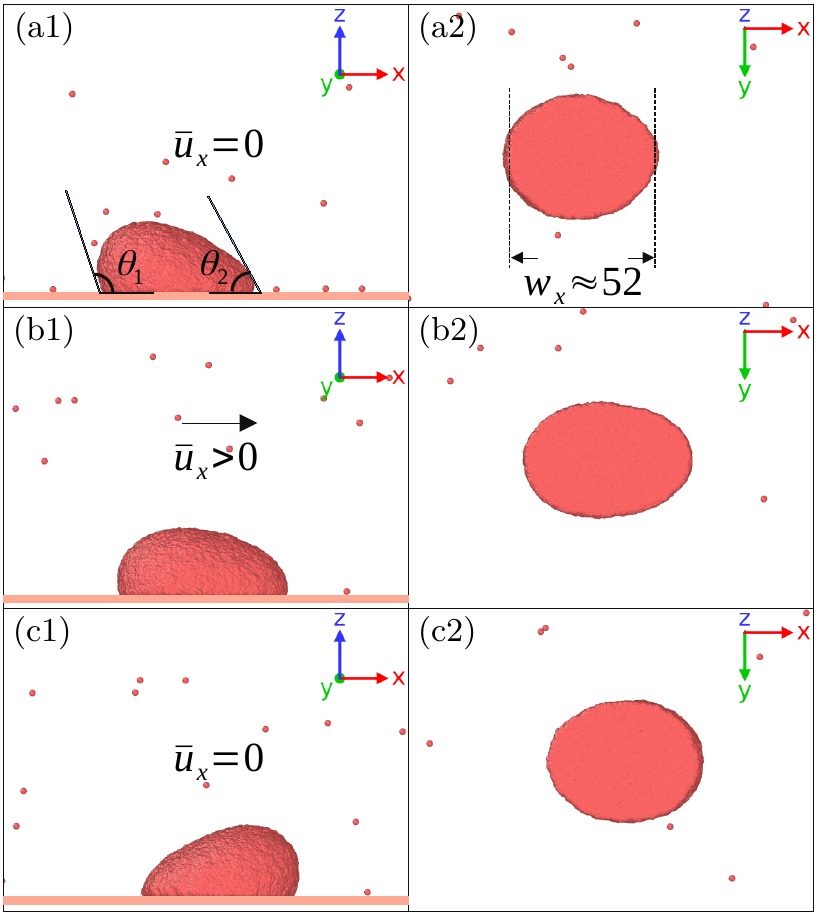}
\caption{Example of Phase I oscillation, with time progressing from (a) to (c). 
$N=20\times10^4$, $\varepsilon_{\rm ws}=2.0$ ($\theta=90^\circ$), 
$A_{\rm{sub}}=900$, $f_{\rm{sub}}=3.75\times10^{-3}$.
$\bar{u}_{x}$ indicates the mean velocity of the droplet along the oscillation direction $x$.
(a1) Side view; (a2) bottom view at the contact surface, 
taken at the same moment as (a1). 
(b1, b2) Side and bottom views taken one-quarter of the oscillation period after (a). 
(c1, c2) Side, bottom views taken one-quarter of the oscillation period after (b).}
\label{fig:phase1_ovito}
\end{figure}

\subsubsection{Phase I oscillation}

Having provided the global picture from our simulations, 
we now proceed with discussing the details
of each scenario. Overall, the simulated
properties in Phase I clearly reflect those observed in
oscillation experiments, where
breakup is avoided.\cite{Gilewicz_Master_Thesis} Hence,
they rather correspond to trivial droplet
oscillation cases, which, however, can be investigated in 
greater detail with MDPD than in experimental measurements.
Figure~\ref{fig:phase1_ovito} illustrates typical snapshots
of Phase I oscillations, which 
are periodic over time without leading to 
droplet breakup. In the
example shown in Figure~\ref{fig:phase1_ovito}, the droplet contains
$N=20\times10^4$ particles, with equilibrium contact angles 
$\theta_1,\theta_2\approx90^{\circ}$ (\textit{i.e.} 
droplet--substrate affinity $\varepsilon_{\rm ws}=2.0$), 
an oscillation amplitude $A_{\rm{sub}}=900$, and frequency 
$f_{\rm{sub}}=3.75\times10^{-3}$. Panels (a) and (c) of 
Figure~\ref{fig:phase1_ovito} correspond to
droplet configurations at times in which the maximum difference
between angles $\theta_1$ and $\theta_2$ is attained.
At these moments, the total velocity (moment) of the 
droplet in the $x$ direction becomes zero. In contrast,
panels (b) of Figure~\ref{fig:phase1_ovito} illustrate a
typical droplet conformation when the droplet has reached
its maximum speed.


\captionsetup[subfigure]{labelformat=empty}
\begin{figure}[htb]

    \centering
    \subfloat[\centering ]{\includegraphics[width=0.47\columnwidth]{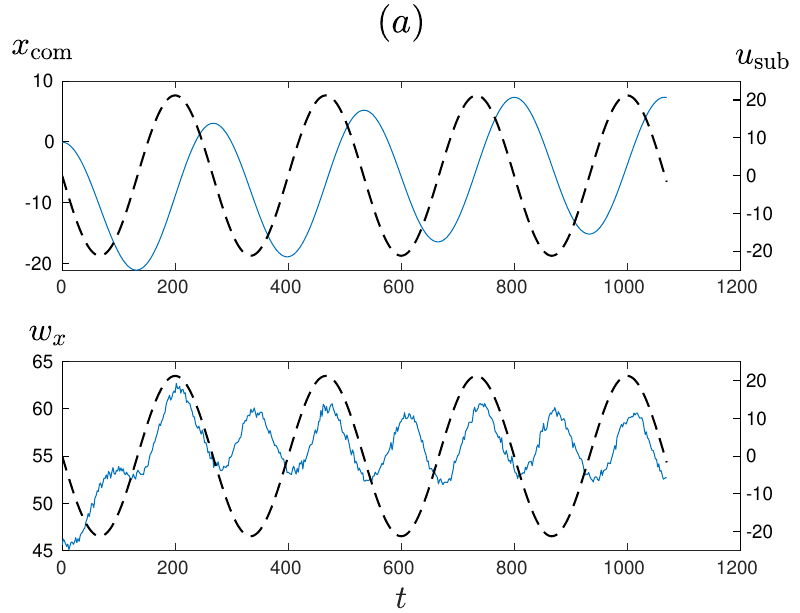} }%
    \qquad
    \subfloat[\centering ]{\includegraphics[width=0.47\columnwidth]{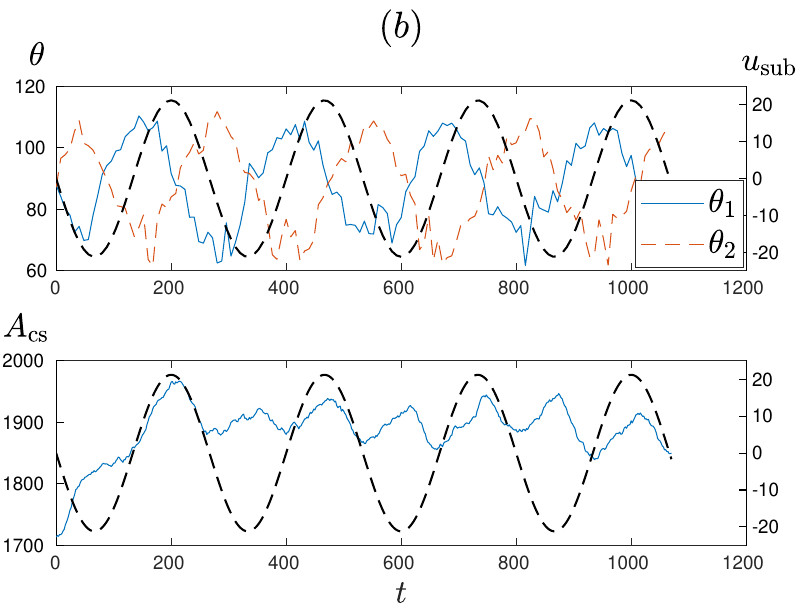} }
    
\caption{Phase I oscillation for the droplet with 
$N=20\times10^4$, $\varepsilon_{\rm ws}=2.0$, $A_{\rm{sub}}=900$, 
$f_{\rm{sub}}=3.75\times10^{-3}$, shown in Figure~\ref{fig:phase1_ovito}. (a) Upper panel: $x_{\rm{com}}$, 
center-of-mass of the droplet; Lower panel: $w_x$, width of the 
droplet in the $x$ direction at the contact surface. 
(b) Upper panel: $\theta_1$ and $\theta_2$  contact angles;
Lower panel: $A_{\rm{cs}}$, contact surface area of the droplet. All
shown in comparison with the substrate velocity $u_{\rm{sub}}$, 
which is presented in the plots by dashed lines.
All plots are obtained from a single simulation trajectory.}
\label{fig:phase1_data}
\end{figure}

The above observations can be quantified by monitoring various
properties, such as the center-of-mass of the droplet, $x_{\rm com}$,
its width, $w_x$, the dynamic contact angles,
$\theta_1$ and $\theta_2$, and the contact surface area, 
$A_{\rm{cs}}$, of the droplet, which are compiled
in Figure~\ref{fig:phase1_data}. These properties are
compared to the periodic motion of the substrate, in this
case, its instantaneous velocity, $u_{\rm sub}$.
In particular, we can observe that $x_{\rm com}$ follows the
periodic pattern of the substrate vibrations with 
a time lag of a quarter of the oscillation period, 
which persists over time during the simulation
(Figure~\ref{fig:phase1_data}(a)).
On the contrary, the width of the droplet $w_x$ (the width in the $x$ direction at the contact surface,
placed at $z=10$, as shown in Figure~\ref{fig:phase1_ovito}(a2)) 
closely follows the periodicity of the substrate velocity, 
$u_{\rm sub}$ (Figure~\ref{fig:phase1_data}(a)). In this case, 
the droplet obtains its maximum width in the
$x$ direction when the substrate velocity, $u_{\rm sub}$,
is also at its maximum, and the droplet relaxes towards
its new forced equilibrium state (Panels (b) of
Figure~\ref{fig:phase1_ovito}), resembling that of a droplet on 
a static substrate when its center-of-mass
velocity is zero. This state is a little more spread out --- the
droplet never recovers the initial width of 46, oscillating 
around the natural width of 55 as long as the vibration remains on.
In contrast, the droplet obtains its minimum width
when the substrate velocity becomes instantaneously 
zero. At this point, the difference between angles
$\theta_1$ and $\theta_2$ becomes maximum
(Panels (a) and (c) of
Figure~\ref{fig:phase1_ovito}).
In particular, the contact angles of the droplet exhibit a stable periodic change, 
with a phase shift of a quarter of the oscillation period relative to 
the substrate velocity (Figure~\ref{fig:phase1_data}(b)). 
For the droplet on a static substrate with a contact angle of
about $90^{\circ}$, the receding angle is around $62^{\circ}$
and the advancing angle is around $110^{\circ}$.
The contact angles $\theta_1$ and $\theta_2$ alternatively 
reach the receding and advancing angles periodically. 
In this example, with droplet--substrate affinity
$\varepsilon_{\rm ws}=2.0$, 
oscillation amplitude $A_{\rm{sub}}=900$, 
and frequency $f_{\rm{sub}}=3.75\times10^{-3}$, 
both angles return to $90^{\circ}$ once per oscillation period
when the amplitude of the substrate velocity, $u_{\rm{sub}}$, 
reaches its maximum, as shown in the snapshot in 
Figure~\ref{fig:phase1_ovito}(b).
This periodic pattern of contact 
angles can vary depending on the oscillation amplitude 
and frequency, as well as the wettability,
determined by $\varepsilon_{\rm ws}$. 
The contact surface area, $A_{\rm{cs}}$, also shows 
periodic behavior, matching the changes in droplet 
width at the contact surface.

\begin{figure}[htb]
\includegraphics[width=1\columnwidth]{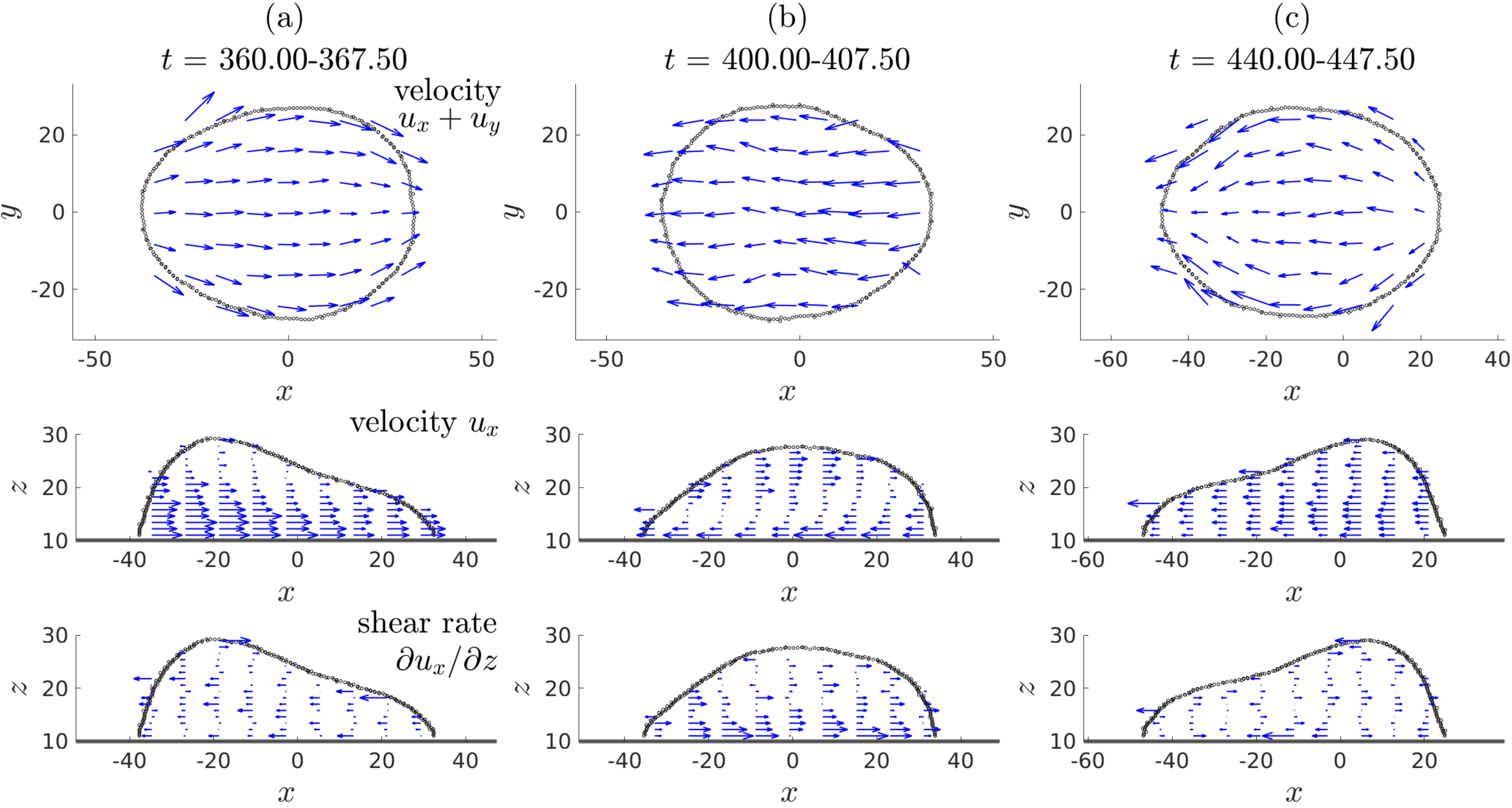}
\caption{Phase I oscillation for the droplet with
$N=20\times10^4$, $\varepsilon_{\rm ws}=3.0$ ($\theta=50^\circ$), $A_{\rm{sub}}=225$, 
$f_{\rm{sub}}=7.5\times10^{-3}$. 
Upper panels: time-averaged velocity field at the contact surface
for the time intervals 
(a) $t=360\sim367.5$, 
(b) $t=400\sim407.5$, 
(c) $t=440\sim447.5$. 
Middle panels: time-averaged velocity field in the $x$ direction
at the cross-section at $y=0$, over the same time intervals 
as the top row. 
Lower panels: time-averaged shear rate, 
$\dot{\gamma}=\partial u_{x}/\partial z$, 
obtained from averaging the consecutive
shear rates. Each velocity field is time-averaged over four 
consecutive frames, with time-interval $t_{\rm{int}}=7.5$. 
}
\label{fig:phase1_vxy}
\end{figure}

We have carefully analyzed the velocity profiles
of the droplets by examining particle-velocity data for
cross-sections of the droplet in various planes at each time
and present representative cases here. 
In Figure~\ref{fig:phase1_vxy}, which shows a different
wetter case in phase I than Figures~\ref{fig:phase1_ovito} 
and~\ref{fig:phase1_data}, now having
$N=20\times10^4$, 
$\varepsilon_{\rm ws}=3.0$ 
(equilibrium contact angles $\sim50^{\circ}$), $A_{\rm{sub}}=225$,
$f_{\rm{sub}}=7.5\times10^{-3}$, such velocity profiles are
plotted for cross-sections in
the $x-y$ and $x-z$ planes. After evaluating the different
profiles, we identified that the most representative profiles
are those on the $x-y$ and $x-z$ planes 
for the velocity components ($u_x$, $u_y$) and $u_x$, 
respectively, as presented in Figure~\ref{fig:phase1_vxy}.
In particular, the $x-y$ cross-section is taken
close to the contact surface and spans the entire droplet, 
while the $x-z$ cross-section includes the entire droplet along 
the $x$ direction. In addition to the velocity profiles,
we have also determined the gradient of the velocity
$u_x$ with respect to the $z$ coordinate, which 
reflects the shear rate $\dot{\gamma}$ in the $z$ direction. 
This analysis allows us to further understand the internal 
velocity distribution and shear dynamics within the droplet.

\begin{figure}[htb]
    \centering
    \subfloat[\centering ]{\includegraphics[width=0.47\columnwidth]{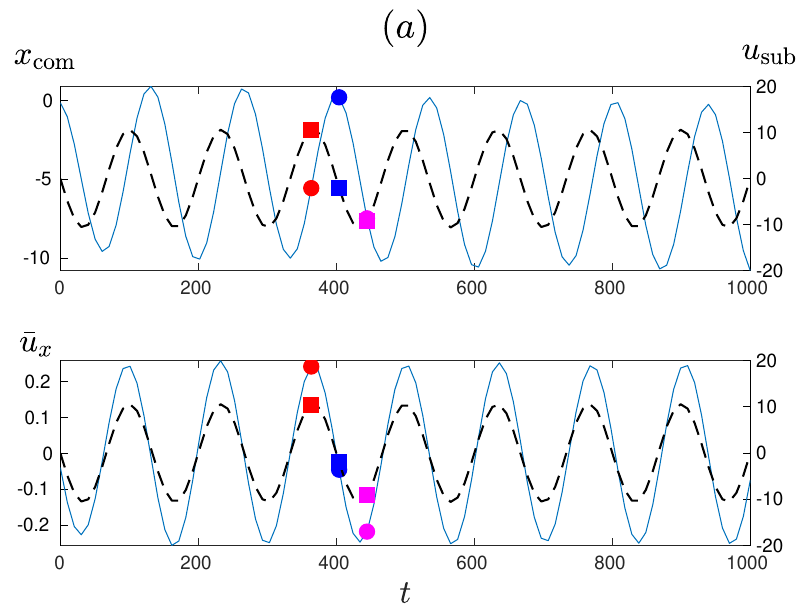} }%
    \qquad
    \subfloat[\centering ]{\includegraphics[width=0.47\columnwidth]{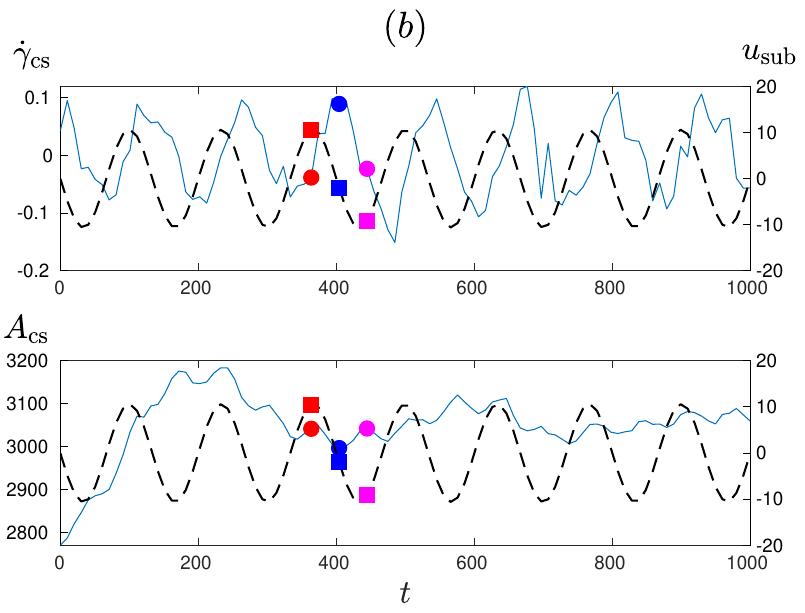} }
    \caption{Time sequence data for the droplet shown in Figure~\ref{fig:phase1_vxy}. 
    (a) Upper panel: center-of-mass position of the droplet; 
    Lower panel: averaged velocity of the droplet.
    (b) Upper panel: average shear rate at the contact surface, $\dot{\gamma}_{\rm{cs}}$; 
    Lower panel: droplet contact surface area. 
    Square markers  represent the substrate velocities,
    while dot markers indicate the measured quantities. 
    Pairs of square and dot markers with the same color 
    correspond to the same time intervals
    in Figure~\ref{fig:phase1_vxy}.
    Substrate velocity, $u_{\rm sub}$, is represented by dashed lines.}%
\label{fig:figscalar_phase1}
\end{figure}

The results for Phase I droplet oscillations, 
illustrated in Figure~\ref{fig:phase1_vxy} are taken at the
specific times indicated by square and dot markers
in Figure~\ref{fig:figscalar_phase1}. The time trace in the 
latter figure (lower panel in subplot (a))
reveals that the mean velocity of the droplet, $\bar{u}_{x}$,
here in this stronger wetting case, would follow that of the 
substrate velocity, 
regardless of the exact moment during the oscillation cycle 
at which these data are recorded. 
In contrast to the previous discussed Phase I system
shown in Figure \ref{fig:phase1_ovito} and \ref{fig:phase1_data}, 
where the droplet attains a symmetric shape similar to that of
a static droplet when both the droplet and substrate velocities 
reach their maximums, in the current case shown in
Figure \ref{fig:phase1_vxy} and \ref{fig:figscalar_phase1}, 
the droplet exhibits a symmetric shape resembling the static droplet
when the velocities of both the droplet and substrate are around 
zero. In addition, the flow field is not uniform. 
The velocities in the $x-y$ plane near the center of the 
droplet at $y=0$ tend to follow the direction of the
vibration,
with particle velocities aligning with the vibration direction.
In contrast, particle velocities along the contact line tend to 
follow the outline of the droplet,
especially when the droplet reaches one of its extrema during 
the oscillation cycle (Figure~\ref{fig:phase1_vxy}(a) and (c)).

\captionsetup[subfigure]{labelformat=empty}

In the case of the velocity field in the $x-z$ plane 
and the shear rate,
we observe periodic patterns with the largest
shear rate as measured in the middle region ($y=0$) of the
droplet at the contact surface, $\dot{\gamma}_{\rm{cs}}$, 
occurring when both the substrate and droplet 
velocities are close to zero (Figure~\ref{fig:figscalar_phase1}(b)). 
This corresponds to the droplet being at the oscillation extrema. 
In this case, the shear rate peaks,
due to the changes in momentum as the
droplet follows the vibrating substrate moving 
towards its center of oscillation. 
This moment is reflected in the symmetric shape of the droplet (Figure~\ref{fig:phase1_vxy}(b)). 
Moreover, the shear rate becomes zero when both the droplet
and the substrate attain their maximum kinetic
energy. In this case, the stresses arising from 
differences in the velocities of the particles
are minimum near the contact surface, and the droplet obtains
an asymmetric shape (Figure~\ref{fig:phase1_vxy}(a) and (c)).
We also observe in the case of Phase I droplets, that the periodic
oscillations attain a steady periodic motion after a few
oscillations. 
This shift to a steady motion is best seen in this example in the
contact surface area $A_{\rm cs}$ (Figure~\ref{fig:figscalar_phase1}(b)). 
The area requires a larger time scale than the other observables
to reflect a steady state. In this particular example, this
time is about four complete oscillations, and a 
somewhat periodic pattern then appears in the
data regarding the contact area with a shorter
and a longer time scale. 
Similar patterns have been observed for Phase I 
oscillations regardless of droplet size and
substrate wettability (see e.g. width and area in Figure~\ref{fig:phase1_data}, where the steady oscillation is attained faster), suggesting it is a general feature. 

\subsubsection{Phase II oscillation}

\begin{figure}[htb!]
\includegraphics[width=0.6\columnwidth]{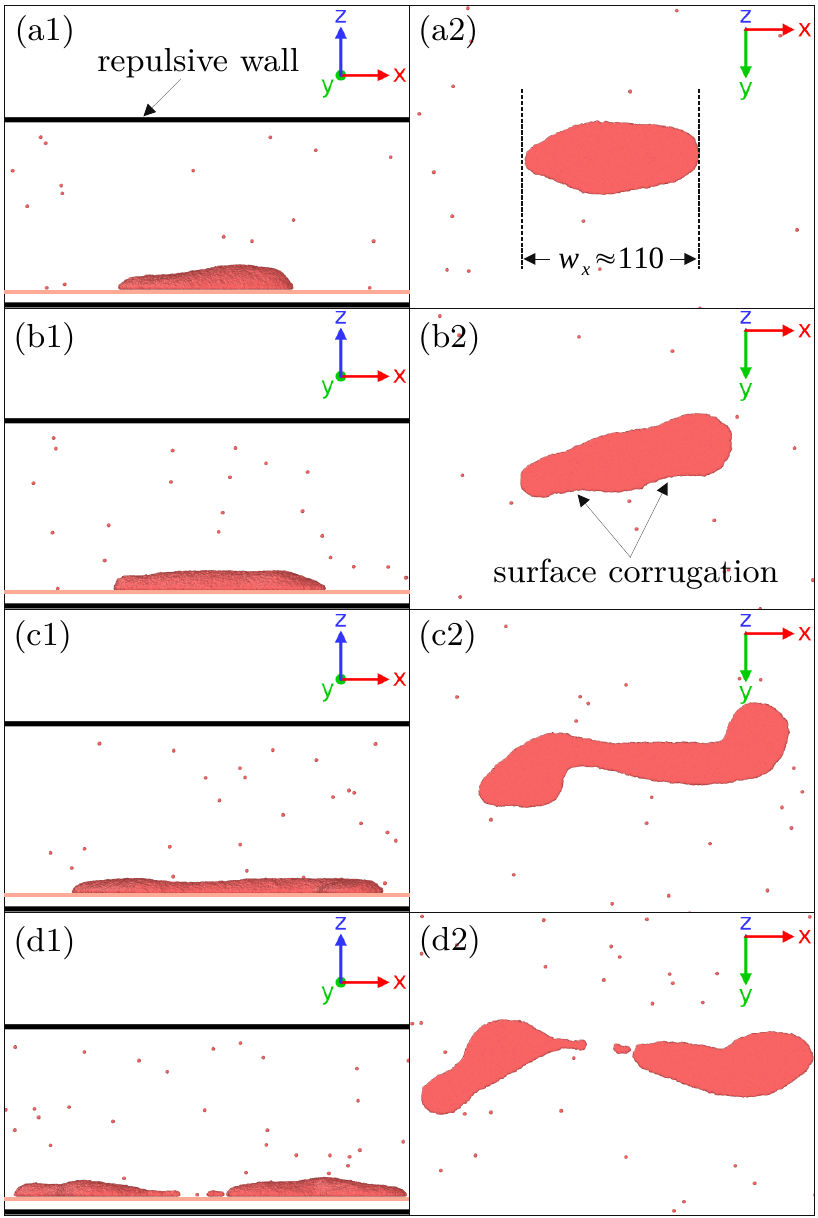}
\caption{Example of Phase II oscillation, with time progressing
from (a) to (d). $N=20\times10^4$,
$\varepsilon_{\rm ws}=3.0$ ($\theta=50^{\circ}$), 
$A_{\rm{sub}}=450$, $f_{\rm{sub}}=7.5\times10^{-3}$.
The simulation box boundary is indicated. 
(a1) Side view; 
(a2) bottom view at the contact surface, at the same moment as (a1),
showing the droplet in a symmetric shape. 
(b1, b2) Internal rotation and stretching develops.
(c1, c2) The droplet stretches further and deviates
significantly from a spherical-cap shape. 
(d1, d2) Droplet break-up occurs. Panels correspond to times 
$t=837, 1532, 2837, 3020$ in Figure~\ref{fig:phase2}. }
\label{fig:phase2_ovito}
\end{figure}

When the substrate becomes more wettable, instabilities
find it easier to arise, leading to the eventual 
breakup of the droplet
for a certain range of substrate vibration amplitude
and frequency (see Figure~\ref{fig:phase_large_droplet}
for the exact values for different contact-angle cases).
We will refer to the scenario in which this occurs after some build-up as Phase II oscillation.
A typical Phase II droplet oscillation is presented
in Figure~\ref{fig:phase2_ovito}. 
In particular, the droplet initially exhibits oscillations
as in Phase I. However, larger conformation changes are recorded
during the simulation and stretching ensues with a rate that
is not tied to the oscillation frequency. 
An internal rotation then develops within the droplet, 
resulting in the surface corrugation that rotates along the contact line, as shown in panel (b) of
Figure~\ref{fig:phase2_ovito}. This phenomenon will be discussed in detail later. Consequently, the droplet assumes an asymmetric shape (in the sense that the droplet 
considerably deviates from 
a spherical-cap shape), as depicted in panel (c) of
Figure~\ref{fig:phase2_ovito}. As a result, 
the left and right domains of the droplet rotate
with a certain momentum, causing these domains to move apart
and eventually leading to the break-up of the droplet
as shown in panel (d) of Figure~\ref{fig:phase2_ovito}. 
Moreover, the formation of satellite droplets is observed
through a mechanism that has been described in
a previous study.\cite{carnevale2023}


\begin{figure}[htb]


    \includegraphics[width=0.6\columnwidth]{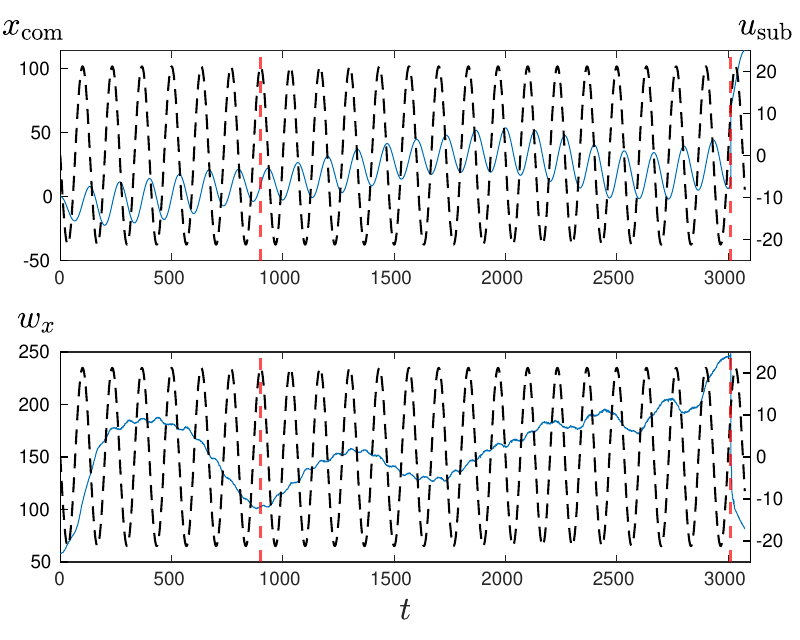}

\caption{Phase II oscillation of the same droplet as 
in Figure~\ref{fig:phase2_ovito}. 
(Upper panel) $x_{\rm com}$: center-of-mass of the droplet; 
(Lower panel) $w_x$: width of the droplet in the $x$ direction 
at the contact surface, 
in comparison with the substrate velocity, $u_{\rm sub}$
(indicated by the dashed line in each graph). 
The asymmetric shape of the droplet, begins to appear around 
$t\sim900$ and develops to the rotated asymmetric shape as 
shown in Figure~\ref{fig:phase2_ovito}(b). 
Droplet breakup occurs at $t=3012$, after which the separated 
droplets oscillate apart as shown in
Figure~\ref{fig:phase2_ovito}(d).
Only the larger droplet is tracked after the breakup.
Both events are marked by red dashed vertical lines.
Plots are obtained from a single trajectory.}
\label{fig:phase2}
\end{figure}

To gain further insights into the mechanisms of Phase II
oscillations, as in the case of Phase I, we have carefully
examined various parameters of the droplets and made
comparison with the Phase I oscillation. 
Figure~\ref{fig:phase2} presents the center-of-mass
oscillations of the droplet in the $x$ direction, $x_{\rm com}$,
and the droplet width in the direction along the oscillation
at the contact surface, $w_x$ (see Figure~\ref{fig:phase2_ovito}(a)
for the definition of the droplet width), up to the point of droplet breakup. They are compared
to the substrate velocity, $u_{\rm{sub}}$. 
In Phase II, $x_{\rm com}$ appears to have a phase shift
of one-quarter relative to $u_{\rm{sub}}$ during the repeated 
vibration cycles, as also seen in Phase I. 
However, a more significant drift in the position $x_{\rm com}$ 
of the droplet is observed in comparison with Phase I. 
This difference in behavior between $x_{\rm com}$ 
and $u_{\rm sub}$ persists 
over the entire oscillation period. 
Two key distinctions between Phase I and Phase II oscillations 
are the emergence of significant out-of-phase stretching and 
the development of asymmetric internal rotation within
the droplet (shown in later figures).
Phase II oscillations are observed to occur under 
moderate oscillation amplitudes and frequencies 
(Figure\ref{fig:phase_large_droplet}),
where the droplet becomes elongated without immediate breakup.
While in Phase I the droplet width, $w_x$, closely follows the 
oscillation pattern of the substrate, in Phase II, $w_x$
follows a much longer period
than the substrate vibrations. This pattern maintains until the 
droplet becomes highly asymmetric,
at which point a non-monotonic change in $w_x$ occurs 
(around $t\sim1700$ in Figure~\ref{fig:phase2}), 
ultimately leading to the droplet breakup.
Moreover, the elongation of $w_x$ is significantly greater
in Phase II. 
For instance, in the Phase I example (Figure~\ref{fig:phase1_data}),
the maximum elongation is approximately $\sim40\%$ 
longer than the static droplet's width,
whereas in Phase II (Figure~\ref{fig:phase2}), 
the maximum elongation reaches about $\sim200\%$. 
This extreme elongation decouples different sections of the droplet, making them susceptible to
independent fluctuations that intensify over time and amplify the droplet’s asymmetry, as shown in panel (b) and (c) of Figure~\ref{fig:phase2_ovito}. 
The oscillating droplet eventually becomes highly asymmetric,
resulting in breakups as shown in panel 
(d) of Figure~\ref{fig:phase2_ovito}.
The increasing asymmetry in the shape of the droplet is evident from the non-monotonic behavior of $w_x$, and the contact surface area $A_{\rm{cs}}$, over time. 

\begin{figure}[htb]
\includegraphics[width=\columnwidth]{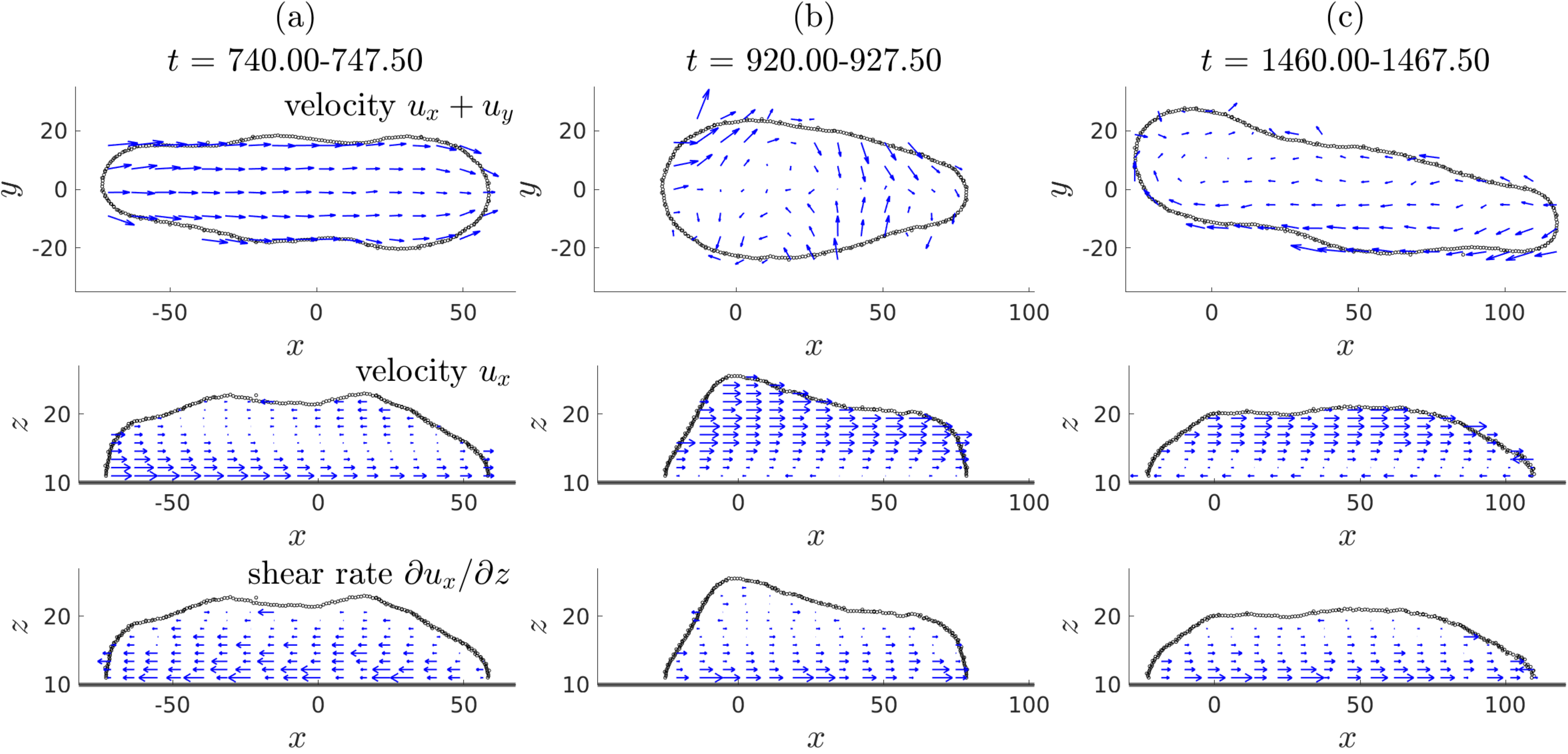}
\caption{Phase II oscillation with identical parameters to 
Figure~\ref{fig:phase2_ovito}, using a different velocity seed. 
Upper panels: time-averaged velocity field at
(a) $t=740\sim747.5$ when the droplet exhibits a elongated shape, 
(b) $t=920\sim927.5$ when velocity-field and rotation
begins to develop, 
(c) $t=1460\sim1467.5$ when the droplet's velocity field
is highly asymmetric. 
Middle panels: time-averaged velocity field in the $x$ direction 
at the cross-section $y=0$, 
taken over the same time intervals as the top row. 
Lower panels: time-averaged shear rate at each stage. 
Each velocity field is time-averaged from four sequential frames.}
\label{fig:phase2_vxy}
\end{figure}

\begin{figure}[htb]

    \centering
    \subfloat[\centering ]{{\includegraphics[width=0.47\columnwidth]{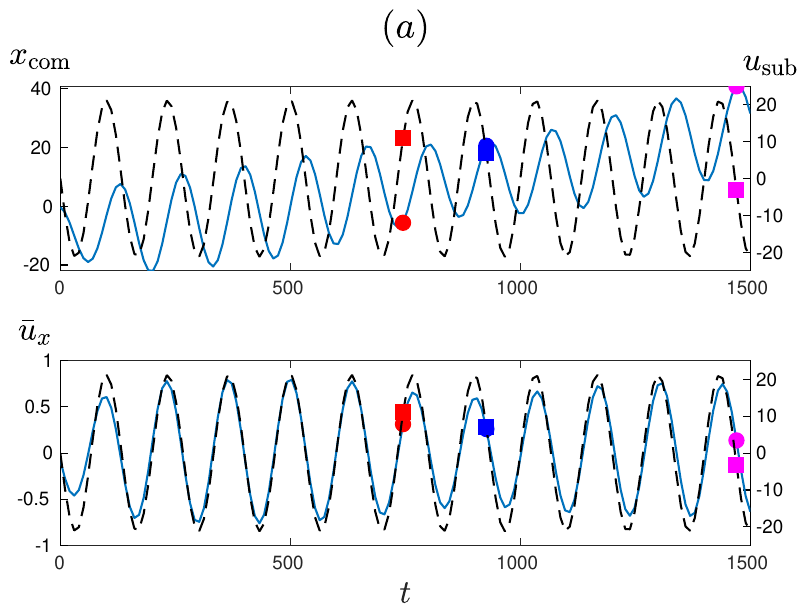} }}%
    \qquad
    \subfloat[\centering ]{{\includegraphics[width=0.47\columnwidth]{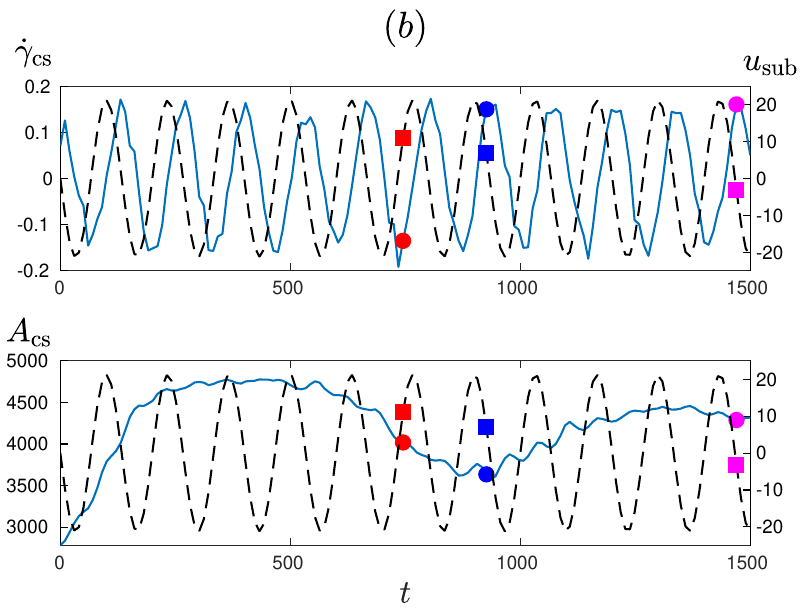} }}%

\caption{(a) Upper panel: center-of-mass position of the droplet from Figure~\ref{fig:phase2_vxy}, with a different seed to Figure~\ref{fig:phase2}; Lower panel: averaged droplet velocity.
(b) Upper panel: average shear rate at the contact surface, 
$\dot{\gamma}_{\rm{cs}}$; 
Lower panel: droplet contact surface area. 
Square markers represent substrate velocities,
while dot markers indicate measured quantities.
Pairs of square and dot markers with the same color 
correspond to matching time intervals in 
Figure~\ref{fig:phase2_vxy}. Substrate velocity $u_{\rm sub}$ 
is represented by dashed lines.
}
\label{fig:figscalar_phase2}
\end{figure}

To explore closer the velocities of the droplet particles 
at the liquid--solid interface,  here, 
we revisit Phase II oscillation 
with parameters identical to those in Figure~\ref{fig:phase2_ovito}
and \ref{fig:phase2},
but with a different initial velocity seed and 
a shorter simulation duration before breakup, as illustrated 
in Figure~\ref{fig:phase2_vxy} and \ref{fig:figscalar_phase2}.
Initially, the velocity field within the elongated droplet aligns
with the substrate's motion  (Figure~\ref{fig:phase2_vxy}(a)).
The droplet then starts to exhibit a rotational
velocity field that provides angular momentum to the 
droplet (Figure~\ref{fig:phase2_vxy}(b)). This results
in a rotational motion of the droplet, reflected in the 
rotation of the surface corrugation, increasing its asymmetry
(Figure~\ref{fig:phase2_vxy}(c)) and ultimately culminating in the breakup of the droplet. The timing of the development 
of the asymmetric velocity field, shape and the
droplet breakup is stochastic. 
In the two cases illustrated in Figure~\ref{fig:phase2_ovito}
and Figure~\ref{fig:phase2_vxy}, the same oscillation parameters 
with different initial velocity seeds yield distinct asymmetric 
droplet shapes and different breakup times. 
However, the oscillation of the droplet width in 
Figure~\ref{fig:phase2} and the contact area in 
Figure~\ref{fig:figscalar_phase2} follows the same
long-time oscillation.
Moreover, the velocity profiles in 
Figure~\ref{fig:phase2_vxy} and the corresponding
shear rates at the particular times shown here, 
indicate that there is a significant difference in
the magnitude of the velocity between the particles
located at the bottom of the droplet and those
closer to the apex. This clearly shows that the upper particles 
cannot fully synchronize with the vibration induced by the 
substrate. Consequently, 
we reason that a sufficiently large thermal
random fluctuation, particularly near the apex,
can result in the asymmetry of the velocity field
observed in Figure~\ref{fig:phase2_vxy}, 
leading to the droplet's rotational motion.

From both Phase II simulations, we observe that the center-of-mass
of the droplet, $x_{\rm com}$, follows the periodicity
of the substrate and droplet velocity, but
with a phase lag of one-quarter of the oscillation period (Figure~\ref{fig:phase2} and ~\ref{fig:figscalar_phase2}(a)).
However, this alignment only persists until the time
that the breakup occurs (Figure~\ref{fig:phase2}). 
Prior to the breakup, $x_{\rm com}$, $\bar{u}_{\rm x}$ 
and $\dot{\gamma}_{\rm cs}$ exhibit periodic patterns (Figure~\ref{fig:figscalar_phase2}) similar to
those in Phase I oscillation. The oscillation of $w_x$ in
Figure~\ref{fig:phase2} and $A_{\rm cs}$ in Figure~\ref{fig:figscalar_phase2} follows the same long-time
oscillation with significant amplitudes.
These observations suggest that the substrate vibrations introduce a systematic driving of the droplet width, progressively increasing asymmetry, and
ultimately leading to oscillations that cause 
the droplet to split apart.

\subsubsection{Phase III Oscillation}

\begin{figure}[htb]
\includegraphics[width=0.6\columnwidth]{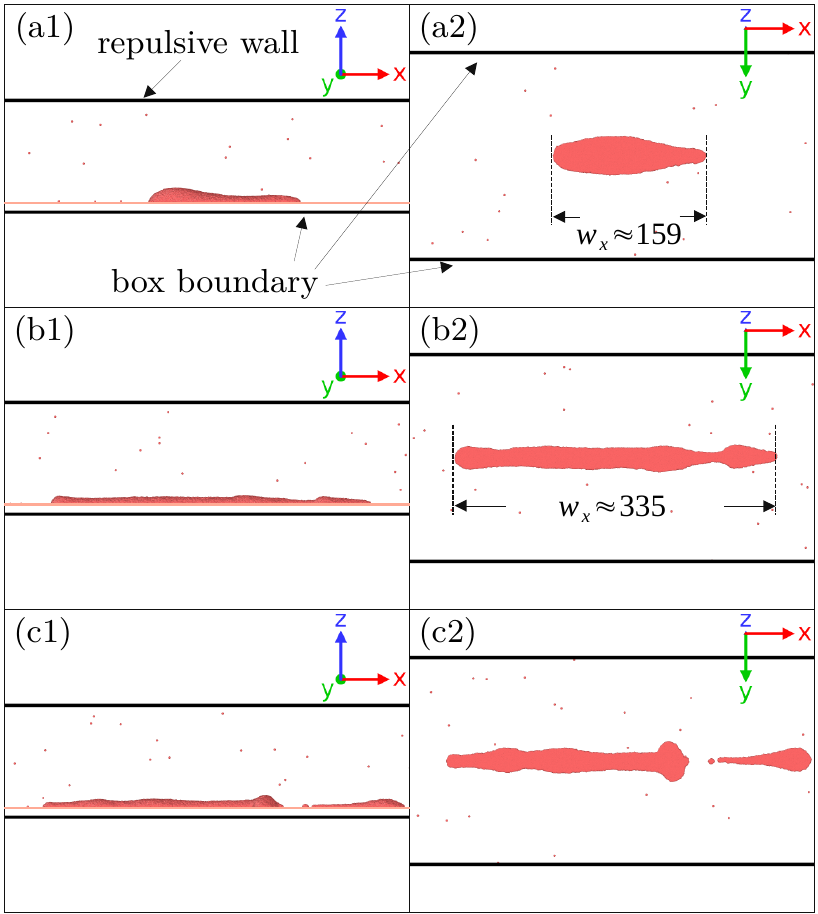}
\caption{Example of phase III oscillation with parameters: 
$N=20\times10^4$, $\varepsilon_{\rm ws}=3.0$ 
($\theta=50^{\circ}$), $A_{\rm{sub}}=675$, $f_{\rm{sub}}=7.5\times10^{-3}$.
(a1) side view; (a2) bottom view at the contact surface, 
taken at the same moment as (a1), 
showing the elongated configuration of the droplet. 
(b1,2) droplet is highly elongated. 
(c1,2) droplet undergoes break-up. Panels correspond to 
times $t=$~100, 350, 450 in Figure~\ref{fig:phase3}.}
\label{fig:phase3_ovito}
\end{figure}

A typical scenario of Phase III oscillation behavior is
demonstrated in Figure~\ref{fig:phase3_ovito}. In the example presented here,  the droplet is subjected to the same oscillation frequency, $f_{\rm{sub}}=7.5\times10^{-3}$, as in the previous Phase II example. However, by increasing the oscillation amplitude to $A_{\rm{sub}}=675$, the droplet transitions to Phase III oscillation. In this phase, the droplet rapidly elongates due to a high
shear rate that eventually leads to the droplet breakup.


\begin{figure}[htb]


    \includegraphics[width=0.6\columnwidth]{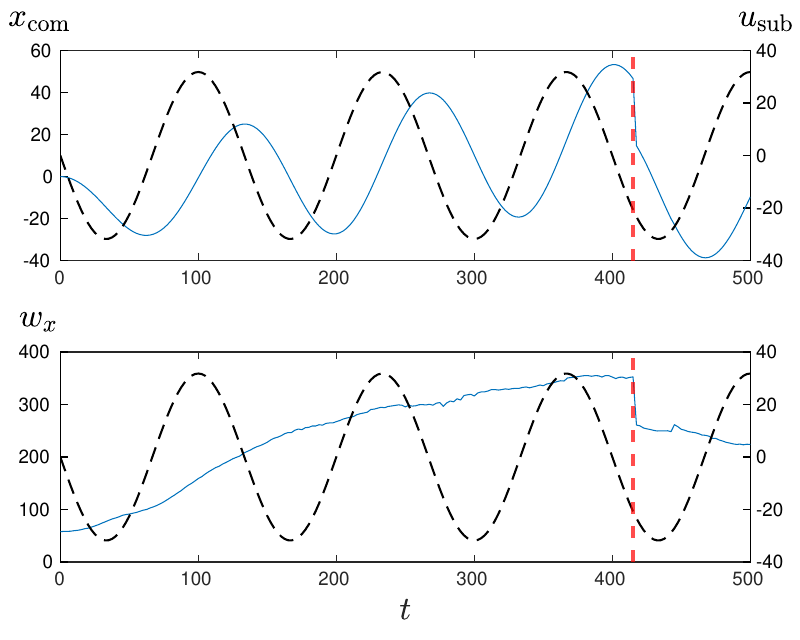}

\caption{Phase III oscillation of the same droplet as shown 
in Figure\ref{fig:phase3_ovito}. 
(Upper panel) $x_{\rm{com}}$: center-of-mass of the droplet; 
(Lower panel) $w_x$: width of the droplet in the $x$ direction at the
contact surface, compared with the substrate velocity 
(dashed lines), $u_{\rm{sub}}$.
The red vertical dashed line at $t=412$ marks the 
point at which the droplet undergoes breakup. Only the larger droplet is tracked after the breakup.
}
\label{fig:phase3}
\end{figure}

\begin{figure}[htb]
\includegraphics[width=\columnwidth]{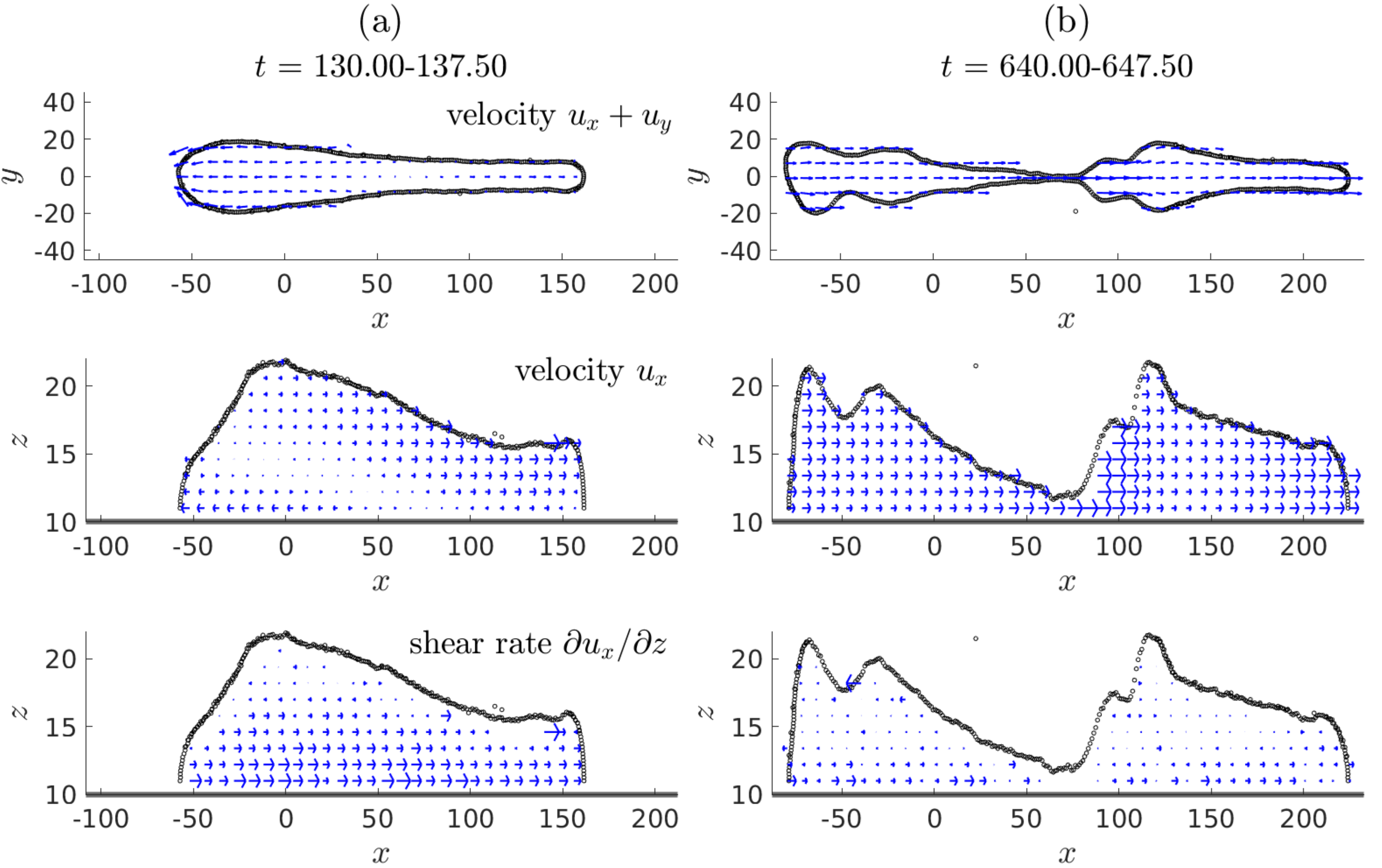}
\caption{Phase III oscillation with identical parameters 
to Figure~\ref{fig:phase3_ovito}, using a different velocity seed.
Upper panels: time-averaged velocity field at (a) $t=130\sim137.5$
when the droplet is elongated in shape, and
(b) $t=640\sim647.5$ at the moment before the droplet breakup. 
Middle panels: time-averaged velocity field in the $x$ direction
at the cross-section, over the same time intervals as the top row. 
Lower panels: time-averaged shear rate. Each velocity field is time-averaged from four sequential frames.}
\label{fig:phase3_vxy}
\end{figure}

In Phase III, the center-of-mass of the droplet, $x_{\rm com}$,
follows the periodic motion of the substrate vibration, but with
a continuous drifting that gradually 
increases (Figure~\ref{fig:phase3}).
This drifting might be due to the continuous elongation
of the droplet, which is indicated by monitoring its width,
$w_x$. The width steadily increases until the breakup of the
droplet, in contrast to the variable and non-monotonic 
change observed in the
breakup of Phase II oscillations. 
After the breakup, only the larger droplet
is tracked. The sudden shift in
$x_{\rm com}$
that is seen when the breakup occurs comes from omission 
of the smaller droplet from the cluster used to calculate 
quantities.

\begin{figure}[htb]

    \centering
    \subfloat[\centering ]{{\includegraphics[width=0.47\columnwidth]{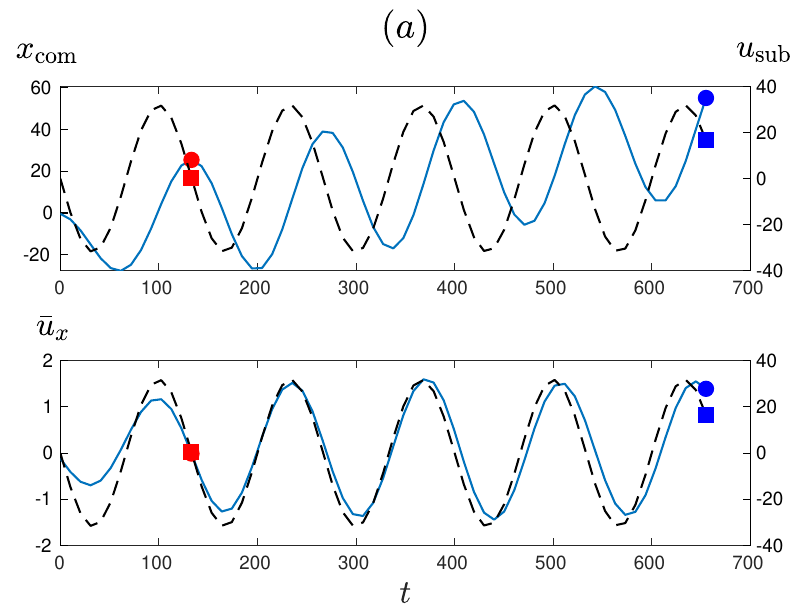} }}%
    \qquad
    \subfloat[\centering ]{{\includegraphics[width=0.47\columnwidth]{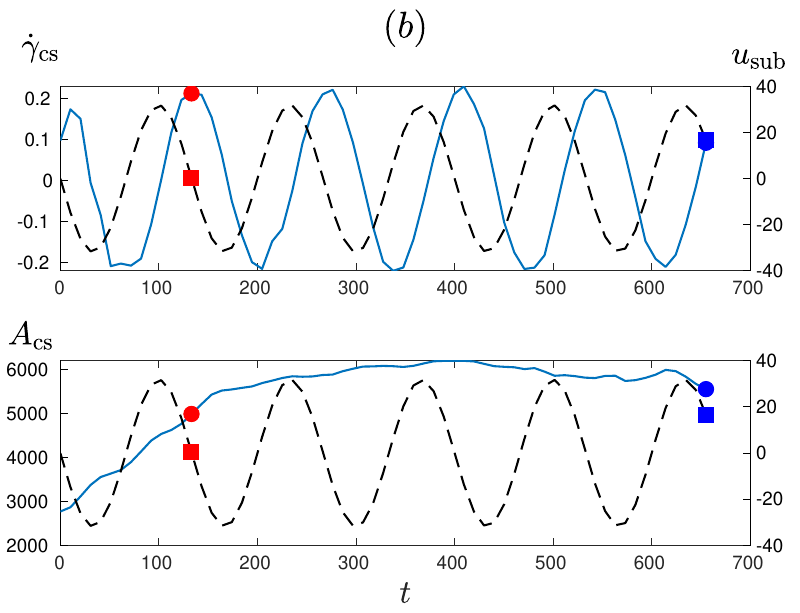} }}%

\caption{(a) Upper panel: center-of-mass position of the droplet described in Figure~\ref{fig:phase3_vxy}; 
Lower panel: averaged droplet velocity. 
(b) Upper panel: average shear rate at the contact surface, 
$\dot{\gamma}_{\rm{cs}}$;
Lower panel: droplet contact surface area. 
Square markers represent substrate velocities, 
while dot markers indicate measured quantities.
Pairs of square and dot markers with the same color
correspond to matching time intervals in Figure~\ref{fig:phase3_vxy}. The blue markers at $t=640\sim647.5$ correspond to the moment before the droplet breakup. Substrate velocity $u_{\rm sub}$ is represented by dashed lines.}
\label{fig:figscalar_phase3}
\end{figure}

By monitoring the velocity profile (Figure~\ref{fig:phase3_vxy}),
we observe that the two domains of the droplet can move in
opposite directions during the elongation stage 
(middle panel of Figure~\ref{fig:phase3_vxy}(a)). 
This opposing motion results in the extreme elongation of the 
droplet during the initial oscillations.
At the time of the breakup,
it is clearly seen that the velocity in the region 
where the two droplets begin to separate differs from the velocity at the centers of the two forming droplets 
(middle panel of Figure~\ref{fig:phase3_vxy}(b)). 
Similar to Phase II, random thermal fluctuations alter the breakup
time. In the two examples shown in Figures~\ref{fig:phase3_ovito} and~\ref{fig:phase3}, 
and Figures~\ref{fig:phase3_vxy} and~\ref{fig:figscalar_phase3}, both examples have the same oscillation
parameters but different initial velocity seeds, 
resulting in different breakup times. 
However, in contrast to the rotation motion observed 
in the case of Phase II, the breakup in
Phase III does not require such rotational dynamics.
This can be better explained by examining the shear profiles 
in the lower panel of Figure~\ref{fig:phase3_vxy}(b),
where we can see that two significant shear regions have developed 
within the droplet, located near the centers of the two forming 
droplets. In contrast, the shear in the region between the two
forming droplets is significantly lower. 
This difference in shear eventually leads to the breakup.

\subsection{Phase transition and droplet dynamics at the contact surface}
\begin{figure}[htb]
\includegraphics[width=0.5\columnwidth]{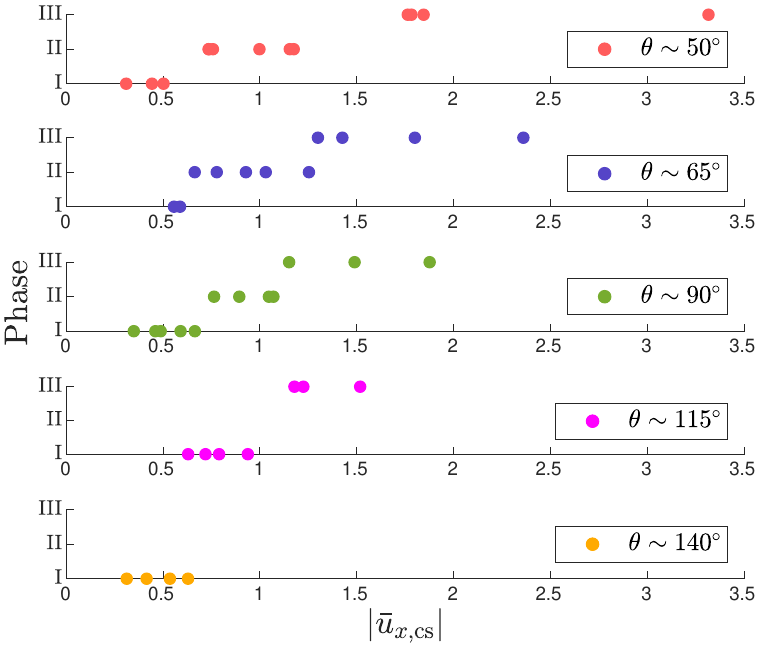}
\caption{Phase versus the magnitude of the average $x$-velocity of the particles at the contact surface, $|\bar{u}_{x,\rm{cs}}|$, for droplets with various equilibrium contact angles $\theta$. Number of particles $N=20\times10^4$.}
\label{fig:phase_CSvelocity}
\end{figure}

In the previous sections, we have examined the role of droplet 
elongation in each phase of droplet oscillation.
In the state diagram shown in Figure~\ref{fig:phase_large_droplet},
it is evident that for the same substrate amplitude, $A_{\rm sub}$,
and frequency, $f_{\rm sub}$ (and thus the same velocity amplitude 
$u_{\rm sub}$), increasing wettability induces a phase transition 
from a lower phase to a higher one. 
The increase in wettability enhances the interaction between the 
droplet particles and the substrate surface,
resulting in higher particle velocities, 
particularly at the contact surface.
Figure~\ref{fig:phase_CSvelocity} illustrates
the magnitude of the average particle velocity in the $x$ direction 
at the contact surface, $|\bar{u}_{x,\rm{cs}}|$, 
and its corresponding phase as observed in the simulation.

Figure \ref{fig:phase_CSvelocity} corresponds to the state diagram for the particle number $N=20\times10^{4}$ in Figure \ref{fig:phase_large_droplet}. This figure clearly demonstrates that the magnitude of the average particle velocity at the contact surface, $|\bar{u}_{x,\rm{cs}}|$, aligns with the oscillation phase of the droplet. In general, across all equilibrium contact angles, $\theta$, the droplet exhibits Phase I oscillation when $|\bar{u}_{x,\rm{cs}}|\ll 1$.
There is a monotonic dependence of phase on $|\bar{u}_{x,\rm{cs}}|$, with low values corresponding to Phase I, high $|\bar{u}_{x,\rm{cs}}|\gg1$ to phase III, and under sufficient wetting $\theta\lesssim 90^{\circ}$, Phase II appears in the vicinity of $|\bar{u}_{x,\rm{cs}}|\sim 1$.
These results, as shown in Figure \ref{fig:phase_CSvelocity}, suggest that the contact surface velocity of the droplet, and consequently the capillary number, plays a critical role in determining the three phases observed in droplet oscillation.

\section{Conclusions}
\label{conclusions}
In this study, we have taken steps to investigate the oscillations of water droplets on horizon-
tally, harmonically, vibrating substrates. The simulation approach followed here offers specific
advantages with regards to experiments and continuum simulation, such as the flawless control
over various conditions and parameters during the simulation experiment and the natural ability
of particle-based simulation to conduct the analysis with a molecular resolution. The main findings of this
study are as follows:

\begin{enumerate}
    \item \textbf{Oscillation Phases:}  
    Droplet oscillations are categorized into three phases based on substrate vibration amplitude and frequency:
    \begin{itemize}
        \item \textbf{Phase I:} Present for both hydrophobic and hydrophilic substrates.
        \item \textbf{Phases II \& III:} More common in hydrophilic substrates, specifically at higher frequencies and amplitudes.  
    \end{itemize}
    We obtained the state diagrams for various droplet sizes.

    \item \textbf{Droplet Dynamics and Shear Stress:}  
    The droplet properties and oscillatory behavior were analyzed with respect to the periodic motion of the substrate. We highlighted the critical role of internal shear stress in droplet breakup.

    \item \textbf{Phase I Dynamics:}  
    Variations in contact angles relative to the substrate velocity in Phase I oscillations at different oscillation amplitudes, frequencies, and droplet--substrate affinities were observed. Future work will explore the contact angle differences and related dynamics in greater detail.

    \item \textbf{Capillary Number Influence:}
    The magnitude of the average particle velocity at the contact surface plays a key role in determining the three oscillation phases, suggesting that the phase behavior is governed by the capillary number of the oscillating droplet.
\end{enumerate}

In summary, this study provides insights into the dynamics of droplet oscillations on harmonically vibrating substrates, hitherto inaccessible to continuum
simulation and experiment.

\section{Supplementary Material}
These are the state diagrams for medium ($N=10\times10^4$) and smaller ($N=5\times10^4$) sized droplets.

\begin{acknowledgments}
This research has been supported by the National 
Science Centre, Poland, under
grant No.\ 2019/34/E/ST3/00232. 
We gratefully acknowledge Polish high-performance computing 
infrastructure PLGrid (HPC Center: ACK Cyfronet AGH) 
for providing computer facilities and support within 
computational grant no. PLG/2023/016608. TB and MK acknowledge 
support from Warsaw University of Technology within the Excellence
Initiative: Research University (IDUB) programme.
\end{acknowledgments}

\bibliography{aipsamp}

\newpage

\begin{figure}[htb]
\includegraphics[width=\columnwidth]{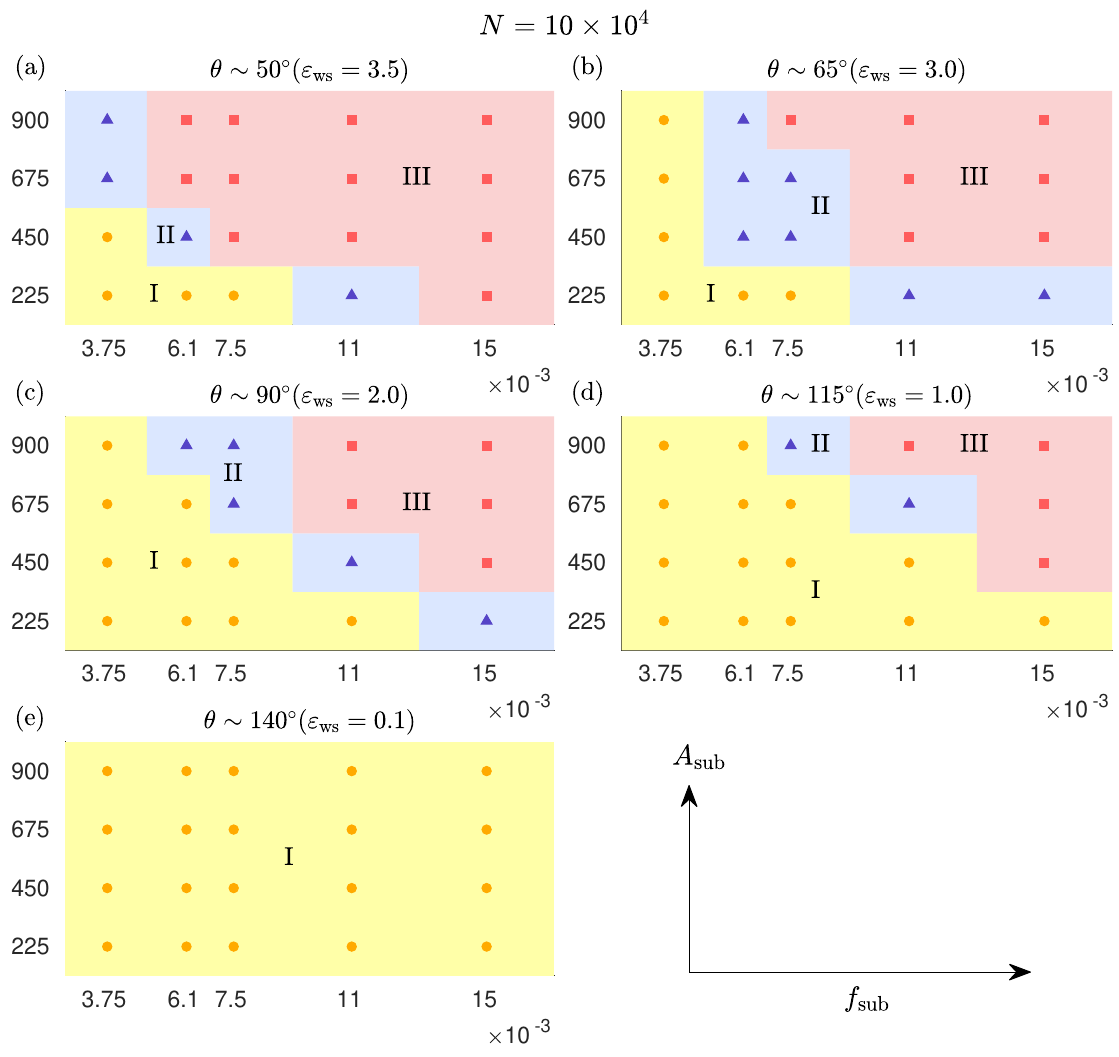}
\caption{State diagram for droplet with number of particles $N=10\times10^4$ at each initial static contact angle $\theta$.}
\label{fig:phase_l_middle_size_droplet}
\end{figure}


\begin{figure}[htb]
\includegraphics[width=\columnwidth]{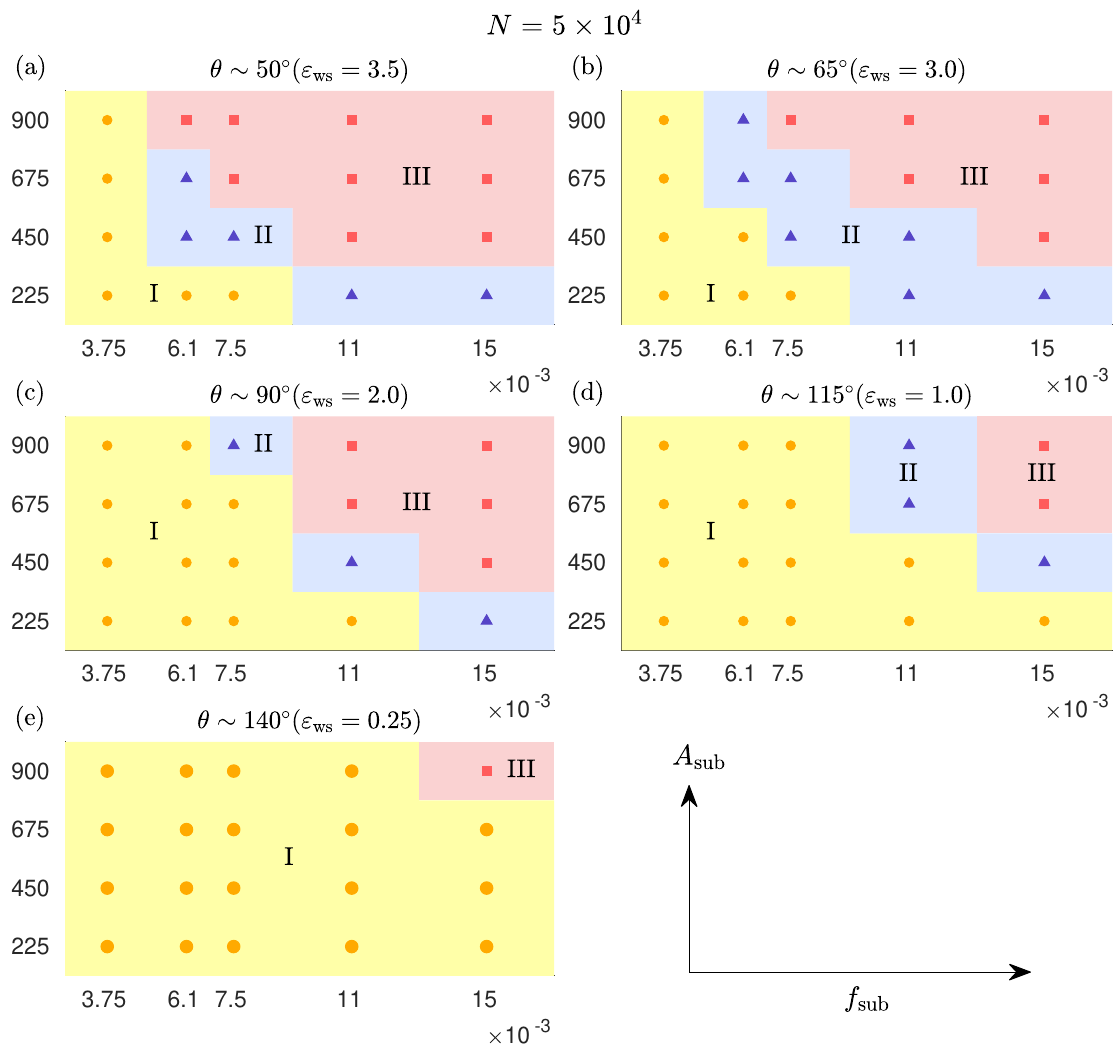}
\caption{State diagram for droplet with number of particles $N=5\times10^4$ at each initial static contact angle $\theta$.}
\label{fig:phase_l_small_droplet}
\end{figure}

\end{document}